\documentclass{elsarticle}

\usepackage[reviewcopy]{adndt}
\usepackage{longtable}

\usepackage{amsmath}

\usepackage{amssymb}

\biboptions{square,sort&compress}
\bibpunct[]{[}{]}{,}{n}{}{;}
\begin{document}

\begin{frontmatter}

\journal{Atomic Data and Nuclear Data Tables}


\title{Discovery of Gallium, Germanium, Lutetium, and Hafnium Isotopes}

\author{J. L. Gross}
\author{M. Thoennessen\corref{cor1}}\ead{thoennessen@nscl.msu.edu}

 \cortext[cor1]{Corresponding author.}

 \address{National Superconducting Cyclotron Laboratory and \\ Department of Physics and Astronomy, Michigan State University, \\ East Lansing, MI 48824, USA}

\begin{abstract}
Currently, twenty-eight gallium, thirty-one germanium, thirty-five lutetium, and thirty-six hafnium isotopes have been observed and the discovery of these isotopes is discussed here. For each isotope a brief synopsis of the first refereed publication, including the production and identification method, is presented.
\end{abstract}

\end{frontmatter}





\newpage
\tableofcontents
\listofDtables

\vskip5pc

\section{Introduction}\label{s:intro}

The discovery of gallium, germanium, lutetium, and hafnium isotopes is discussed as part of the series summarizing the discovery of isotopes, beginning with the cerium isotopes in 2009 \cite{2009Gin01}. Guidelines for assigning credit for discovery are (1) clear identification, either through decay-curves and relationships to other known isotopes, particle or $\gamma$-ray spectra, or unique mass and Z-identification, and (2) publication of the discovery in a refereed journal. The authors and year of the first publication, the laboratory where the isotopes were produced as well as the production and identification methods are discussed. When appropriate, references to conference proceedings, internal reports, and theses are included. When a discovery includes a half-life measurement the measured value is compared to the currently adopted value taken from the NUBASE evaluation \cite{2003Aud01} which is based on the ENSDF database \cite{2008ENS01}. In cases where the reported half-life differed significantly from the adopted half-life (up to approximately a factor of two), we searched the subsequent literature for indications that the measurement was erroneous. If that was not the case we credited the authors with the discovery in spite of the inaccurate half-life.

The first criterion is not clear cut and in many instances debatable. Within the scope of the present project it is not possible to scrutinize each paper for the accuracy of the experimental data as is done for the discovery of elements \cite{1991IUP01}. In some cases an initial tentative assignment is not specifically confirmed in later papers and the first assignment is tacitly accepted by the community. The readers are encouraged to contact the authors if they disagree with an assignment because they are aware of an earlier paper or if they found evidence that the data of the chosen paper were incorrect. Measurements of half-lives of a given element without mass identification are not accepted. This affects mostly isotopes first observed in fission where decay curves of chemically separated elements were measured without the capability to determine their mass. Also the four-parameter measurements (see, for example, Ref. \cite{1970Joh01}) were, in general, not considered because the mass identification was only $\pm$1 mass unit.


The initial literature search was performed using the databases ENSDF \cite{2008ENS01} and NSR \cite{2008NSR01} of the National Nuclear Data Center at Brookhaven National Laboratory. These databases are complete and reliable back to the early 1960's. For earlier references, several editions of the Table of Isotopes were used \cite{1940Liv01,1944Sea01,1948Sea01,1953Hol02,1958Str01,1967Led01}. A good reference for the discovery of the stable isotopes was the second edition of Aston's book ``Mass Spectra and Isotopes'' \cite{1942Ast01}.

\section{Discovery of $^{60-87}$Ga}

Twenty-eight gallium isotopes from A = 60$-$87 have been discovered so far; these include 2 stable ($^{69}$Ga and $^{71}$Ga), 10 neutron-deficient and 16 neutron-rich isotopes. According to the HFB-14 model \cite{2007Gor01}, $^{102}$Ga should be the last odd-odd particle stable neutron-rich nucleus while the odd-even particle stable neutron-rich nuclei should continue through $^{107}$Ga. The proton dripline has most likely been reached at $^{60}$Ga from the non-observance of $^{59}$Ga \cite{2005Sto01}. About 18 isotopes have yet to be discovered corresponding to 39\% of all possible gallium isotopes.

\begin{figure}
	\centering
	\includegraphics[scale=.7]{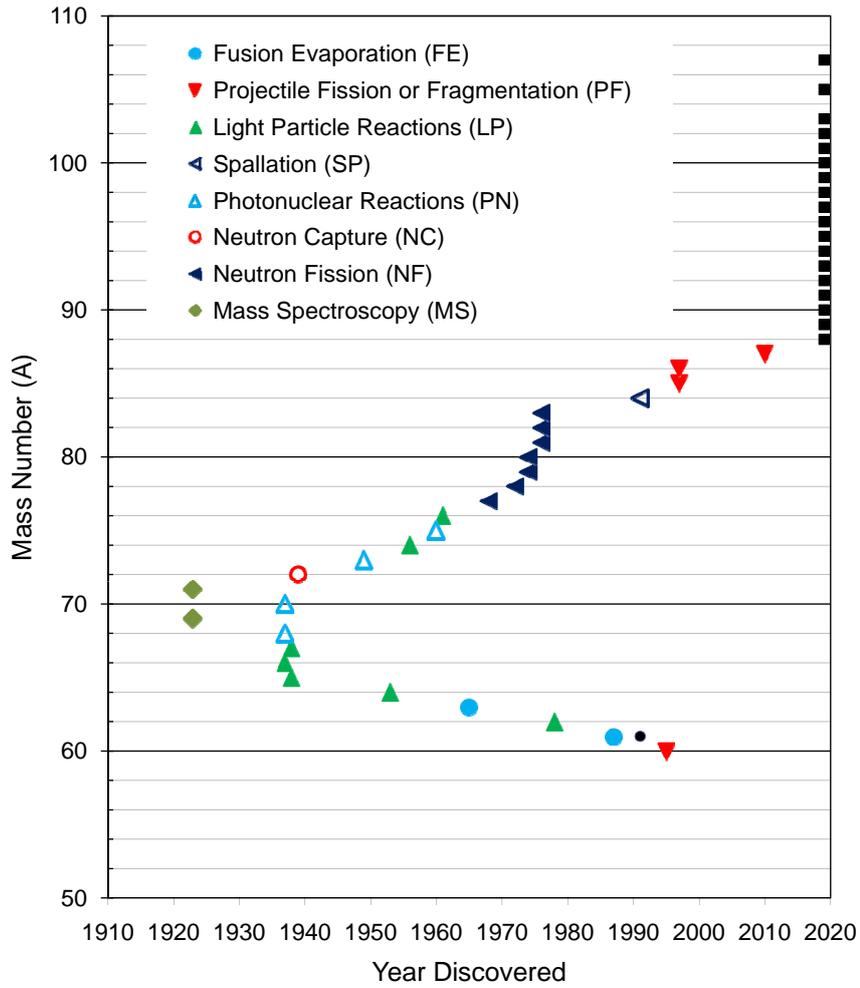}
	\caption{Gallium isotopes as a function of time when they were discovered. The different production methods are indicated. The solid black squares on the right hand side of the plot are isotopes predicted to be bound by the HFB-14 model. The black solid circle indicates the discovery year of the ground state in the case where the first observation was an excited state.}
\label{f:year-gallium}
\end{figure}

Figure \ref{f:year-gallium} summarizes the year of first discovery for all gallium isotopes identified by the method of discovery. The range of isotopes predicted to exist is indicated on the right side of the figure. The radioactive gallium isotopes were produced using fusion evaporation reactions (FE), light-particle reactions (LP), neutron induced fission (NF), neutron-capture reactions (NC), spallation reactions (SP), photo-nuclear reactions (PN) and projectile fragmentation or fission (PF). The stable isotopes were identified using mass spectroscopy (MS). In the following, the discovery of each gallium isotope is discussed in detail and a summary is presented in Table 1.

\subsection*{Stable isotopes $^{69}$Ga and $^{71}$Ga}
In 1923, Aston reported the discovery of stable $^{69}$Ga and $^{71}$Ga in ``Further determinations of the constitution of the elements by the method of accelerated anode rays'' \cite{1923Ast01}. Gallium fluoride was used in the Cavendish mass spectrometer. ``Gallium fluoride made from a specimen of the hydrate kindly provided by Prof.\ Richards, of Harvard University, also gave satisfactory results. Gallium consists of two isotopes, 69 and 71.''

\subsection*{$^{60}$Ga}
In the 1995 article ``New isotopes from $^{78}$Kr fragmentation and the ending point of the astrophysical rapid-proton-capture process'' Blank et al.\ reported the discovery of $^{60}$Ga \cite{1995Bla01}. A 73~MeV/nucleon $^{78}$Kr beam bombarded a nickel target \cite{1995Bla01} at GANIL. $^{60}$Ga was produced via projectile fragmentation and identified with the SISSI/LISE facility by measuring the time-of-flight through the separator and the $\Delta$E-E in a silicon detector telescope. A lower limit for the half-life was established, ``We find clear evidence for $^{60}$Ga, $^{64}$As, $^{69,70}$Kr, and $^{74}$Sr.''

\subsection*{$^{61}$Ga}
In ``Beta-delayed proton decay of $^{61}$Ge'' Hotchkis et al.\ reported the observation of a proton-unbound state of $^{61}$Ga in 1987 \cite{1987Hot01}. A $^{24}$Mg beam of 77--120 MeV energy from the Berkeley 88-in.\ cyclotron bombarded a calcium target. Beta-delayed protons were measured with a silicon telescope in connection with a pulsed beam technique and a helium-jet transport system. ``The 3.11 MeV peak is attributed to the decay of $^{61}$Ge, since its yield is a maximum at 90 MeV, as expected from the ALICE calculations. This peak is identified as the transition from the isobaric analog state in $^{61}$Ga, fed by the superallowed beta decay of $^{61}$Ge, to the ground state of $^{60}$Zn.'' The ground state of $^{61}$Ga was observed four years later \cite{1991Moh01}.

\subsection*{$^{62}$Ga}
Chiba et al.\ identified $^{62}$Ga in their 1978 paper, ``Superallowed fermi beta transitions of $^{62}$Ga'' \cite{1978Chi01}. 36$-$52 MeV protons from the Tokyo synchrocyclotron bombarded a natural zinc target. $^{62}$Ga was produced in the reaction $^{64}$Zn(p,3n) and $\beta$-activities were measured with a plastic scintillator. ``The obtained data points were analyzed with a least squares fitting program by taking into account the following three components: a 20-msec $^{12}$B, an unknown $^{62}$Ga, and a constant long lived background activity. The resulting decay curve is shown in [the figure], and a value of T$_{1/2}$ = 116.4$\pm$1.5 msec was obtained for the half-life of $^{62}$Ga.'' This half-life agrees with the currently accepted value of 116.12(23)~ms.

\subsection*{$^{63}$Ga}
The first observation of $^{63}$Ga was published by Nurmia and Fink in 1965 in ``A new short-lived isotope of gallium'' \cite{1965Nur01}. The Argonne 66-in.\ cyclotron was used to bombard nickel targets with a 66~MeV $^6$Li beam and $^{63}$Ga was formed in the reactions $^{58}$Ni($^6$Li,n) and $^{60}$Ni($^6$Li,3n). Beta-ray decay curves were measured with an end-window proportional counter following chemical separation. ``The half-life is 33$\pm$4~s, and its assignment to Ga$^{63}$ has been established by radiochemical separations, including isolation of the 38.3~m Zn$^{63}$ daughter activity, and by cross bombardments.'' This half-life agrees with the presently adopted value of 32.4(5)~s.

\subsection*{$^{64}$Ga}
``Gallium-64'' by Cohen was the first report of $^{64}$Ga in 1953 \cite{1953Coh01}. Protons from the Oak Ridge 86-in.\ cyclotron bombarded enriched $^{64}$Zn targets and $^{64}$Ga was produced in the (p,n) charge exchange reaction. Gamma-ray spectra were measured with a NaI scintillation spectrometer following chemical separation. ``A new isotope, 2.5-minute Ga$^{64}$, was produced by the (p,n) reaction on Zn$^{64}$ and identified by measurement of the excitation function, by bombardment of separated isotopes, and by chemical separation.'' This half-life agrees with the currently accepted value of 2.627(12)~m. Less than a month later, Crasemann independently reported a half-life of 2.6(1)~min for $^{64}$Ga \cite{1953Cra01}. An earlier 48(2)~min half-life measurement \cite{1938Buc01} had been refuted by \cite{1952Muk01}.

\subsection*{$^{65}$Ga}
In 1938, $^{65}$Ga was reported in ``The capture of orbital electrons by nuclei'' by Alvarez \cite{1938Alv02}. Zinc samples were bombarded with fast deuterons and subsequent emission of X-rays and electrons were measured. Evidence for $^{65}$Ga was given in a footnote: ``The shortest of these four periods in unseparated, activated zinc was a new electron emitting isotope with a half-life of 15~minutes. The X-rays are definitely Zn K-radiation, and since this period is unknown, it might be due to Ga$^{65}$ capturing electrons.'' This half-life agrees with the presently accepted value of 15.2(2)~m. No other measurements of $^{65}$Ga were published until 13 years later when Aten et al.\ credited Alvarez for the first observation of $^{65}$Ga \cite{1952Ate02}.

\subsection*{$^{66}$Ga}
Mann identified $^{66}$Ga in the 1937 paper ``Nuclear transformations produced in copper by alpha-particle bombardment'' \cite{1937Man01}. The Berkeley cyclotron was used to bombard copper targets with 11 MeV $\alpha$-particles. Resulting activities were measured with a quartz fiber electroscope and positrons tracks in a cloud chamber were photographed. ``Activities having half-lives of 1.10$\pm$0.05 hours and 9.2$\pm$0.2 hours have been found to belong, respectively, to the radioactive isotopes of gallium Ga$^{68}$ and Ga$^{66}$.'' This $^{66}$Ga half-life agrees with the currently accepted value of 9.49(3)~h.

\subsection*{$^{67}$Ga}
The first identification of $^{67}$Ga was reported by Alvarez in ``Electron capture and internal conversion of gallium 67'' in 1938 \cite{1938Alv01}. Deuterons bombarded zinc targets and X-rays, $\gamma$ rays, and electrons were measured following chemical separation. ``The activity, of 83 hours half-life, has the chemical properties of Ga, as shown by the solubility of its chloride in ether. It has been assigned to Ga$^{67}$ by Mann, who bombarded Zn with alpha-particles, and separated Ga by chemical means.'' This half-life is close to the presently adopted value of 3.2617(5)~d. The reference to Mann mentioned in the quote reported a 55~h half-life in a conference abstract \cite{1938Man02}. Mann later corrected this value to 79~h \cite{1938Man01}.

\subsection*{$^{68,70}$Ga}
Bothe and Gentner observed $^{68}$Ga and $^{70}$Ga in 1937 in ``Weitere Atomumwandlungen durch $\gamma$-Strahlen'' \cite{1937Bot03}. Lithium-$\gamma$-rays irradiated gallium targets producing $^{68}$Ga and $^{70}$Ga in photo-nuclear reactions. ``Gallium: T$_1$ = 20 min; T$_2$= 60 min. T$_1$ wird auch bei Anlagerung von Neutronen erhalten, geh\"ort also zu Ga$^{70}$, weil dieses zwischen den beiden stabilen Isotopen des Ga liegt. Dann mu\ss\ T$_2$ zu dem neuen Isotop Ga$^{68}$ geh\"oren.'' [Gallium: T$_1$ = 20 min; T$_2$ = 60 min. T$_1$ can also be produced by neutron capture and thus corresponds to Ga$^{70}$ because it is located between the two stable Ga isotopes. Then T$_2$ must be due to the new isotope Ga$^{68}$]. The half-lives of 60~min for $^{68}$Ga and 20~min for $^{70}$Ga agree with the currently adopted values of 67.71(9)~min and 21.14(3)~min, respectively. A 20-min half-life had previously been observed without a mass assignment \cite{1935Ama01}.

\subsection*{$^{72}$Ga}
Sagane identified $^{72}$Ga in 1939 as reported in ``Radioactive isotopes of Cu, Zn, Ga and Ge'' \cite{1939Sag01}. Metallic gallium targets were irradiated by slow and fast neutrons produced by deuteron bombardments of lithium and beryllium from the Berkeley cyclotron. $^{72}$Ga was produced by neutron capture reactions and the resulting activities were measured with a Lauritsen-type quartz fiber electroscope. ``As shown in [the figure] the decay curves obtained with slow neutron bombardments gave only two periods; the well-known 20-min. period and a new 14-hr. period. No trace of the 23-hr. period reported by Fermi and others or by the Michigan group was found... These two points support very well the conclusion that this 14-hr. period should be caused by Ga$^{72}$'' The reported half-life of 14.1(2)~h is included in the calculation of the current value of 14.10(2)~h. Earlier, Amaldi et al.\ reported a 23~h half-life without a mass assignment \cite{1935Ama01}. Subsequently, without measuring it themselves Bothe and Gentner, as well as Mann assigned this activity to $^{72}$Ga \cite{1937Bot04,1937Man01}. Only a few month later Pool et al.\ measured a 22~h period assigning it to $^{72}$Ga \cite{1937Poo01}. In the following year, Livingston reported a 14~h half-life, however, he did not explicitly assigned it to $^{72}$Ga \cite{1938Liv01}.

\subsection*{$^{73}$Ga}
The identification of $^{73}$Ga was reported by Perlman in the 1949 paper ``Yield of some photo-nuclear reactions'' \cite{1949Per01}. A pure germanium oxide target was irradiated with 50 MeV and 100 MeV X-rays and $^{73}$Ga was formed in the photo-nuclear reaction $^{74}$Ge($\gamma$,p). Decay curves and $\beta$-ray spectra were measured following chemical separation. The results were summarized in a table where a half-life of 5~h was listed for $^{73}$Ga. This half-life agrees with the presently adopted value of 4.86(3)~h. A 5.0(5)~h half-life had tentatively been assigned to $^{73}$Ga by Siegel and Glendenin in 1945 as part of the Plutonium Project which was published in the open literature only in 1951 \cite{1951Sie01}.

\subsection*{$^{74}$Ga}
$^{74}$Ga was discovered by Morinaga as reported in ``Radioactive isotopes Cl$^{40}$ and Ga$^{74}$'' in 1956 \cite{1956Mor01}. Germanium targets were irradiated with fast neutrons produced by bombarding a beryllium target with 10 MeV deuterons from the Purdue cyclotron. Gamma- and beta-rays were measured with a NaI scintillator and GM counter, respectively. ``Very many gamma rays with various half-lives were observed after the bombardment, but all could be assigned to some known isotopes produced by fast neutrons on Ge, except for three distinct gamma rays with energies 0.58, 2.3, and 2.6 Mev which decayed with a half-life of about 8 min... Therefore this activity is assigned to Ga$^{74}$.'' This half-life is consistent with the currently adopted value of 8.12(12)~m. Earlier measurements incorrectly assigned half-lives of 6(1)~d \cite{1939Sag01} and 9~d \cite{1941Sag01} to $^{74}$Ga.

\subsection*{$^{75}$Ga}
Morinaga et al.\ discovered $^{75}$Ga in 1960 as reported in ``Three new isotopes, $^{63}$Co, $^{75}$Ga, $^{81}$As'' \cite{1960Mor01}. Metallic germanium was irradiated with 25~MeV bremsstrahlung from the Tohoku betatron and $^{75}$Ga was produced in the photo-nuclear reaction $^{76}$Ge($\gamma$,p). Gamma- and beta-radiation were measured with scintillation spectrometers following chemical separation. ``Besides all the known activities a component which decayed with a half-life of approximately 2 minutes was observed.'' The reported half-life of 2.0(1)~min agrees with the presently accepted value of 126(2)~s.

\subsection*{$^{76}$Ga}
The observation of $^{76}$Ga was described by Takashi et al.\ in the 1961 paper ``Some new activities produced by fast neutron bombardments'' \cite{1961Tak01}. Fast neutrons produced by bombarding graphite targets with 20 MeV deuterons from the Tokyo 160 cm variable energy cyclotron irradiated a metallic germanium sample. Gamma- and beta-ray spectra were measured with NaI(Tl) and plastic scintillators, respectively. ``Since no other product of fast neutron reaction on Ge can give such a high energy electron radiation and the cross section to produce the activity is equal in order of magnitude to that of Ge$^{74}(n,p)$Ga$^{74}$ reaction, the activity is safely assigned to Ga$^{76}$.'' The reported half-life of 32(3)~s agrees with the presently adopted value of 32.6(6)~s.

\subsection*{$^{77}$Ga}
Wish reported the observation of $^{77}$Ga in ``Thermal neutron fission of $^{235}$U: Identification and functional chain yield of 17-sec $^{77}$Ga.'' \cite{1968Wis01}. Thermal neutrons from the Vallecitos Nuclear Test Reactor irradiated a solution of enriched $^{235}$U and Ga(III) carrier in hydrochloric acid. The resulting $\beta$-ray activity was measured with a gas-flow proportional counter following chemical separation. ``The results indicate the presence of $^{77}$Ga and $^{78}$Ga in the fission-product mixture. A plot of the 11.3-h $^{77}$Ge $\beta$-ray activity versus the time of the Ga separation after irradiation is shown in [the figure]. A least-squares fit of the data gave a half-life of 17.1$\pm$1.5~sec for $^{77}$Ga.'' This half-life is close to the presently accepted value of 13.2(2)~s.

\subsection*{$^{78}$Ga}
In the 1972 paper ``Identification of new germanium isotopes in fission: Decay properties and nuclear charge distribution in the A = 78 to 84 mass region'' del Marmol and Fettweis identified $^{78}$Ga \cite{1972del01}. A uranyl nitrate solution of $^{235}$U was irradiated with neutrons from the Mol BR1 graphite reactor. Gamma-ray spectra were recorded with a Ge(Li) detector following chemical separation. ``By using this method a half-life of $4.8\pm1.3$~s is found for $^{78}$Ga; it specifies the estimated value of $\approx$4~s by Wish who counted a $^{78}$Ge-$^{78}$As mixture formed from fission produced gallium, separated different times after the end of irradiation and it confirms the element assignment of a 4.9$\pm$0.2~s half-life obtained through mass separation by the Osiris collaboration.'' This half-life agrees with the currently accepted value of 5.09(5)~s. As stated in the quote, previously Wish only estimated a value for the half-life \cite{1968Wis01}, while the measurement by the Osiris collaboration was only published in a conference proceeding \cite{1970Gra01}.

\subsection*{$^{79,80}$Ga}
$^{79}$Ga and $^{80}$Ga were observed by Grapengiesser et al.\ in the 1974 paper ``Survey of short-lived fission products obtained using the Isotope-Separator-On-Line Facility at Studsvik'' in 1974 \cite{1974Gra01}. The gallium isotopes were produced by neutron induced fission and identified at the OSIRIS isotope-separator online facility. In the first long table, the half-life of $^{79}$Ga is quoted as 3.00(9)~s, which agrees with the currently accepted value of 2.847(3)~s. The authors reference an internal report \cite{1973Rud01} as the source of a half-life of 1.7(2)~s for $^{80}$Ga, which is included in  calculating the currently accepted average value of 1.676(14)~s.

\subsection*{$^{81-83}$Ga}
Rudstam and Lund reported the observation of $^{81}$Ga, $^{82}$Ga and $^{83}$Ga in ``Delayed-neutron activities produced in fission: mass range 79-98'' in 1976 \cite{1976Rud01}. $^{235}$U targets were irradiated with neutrons from the Studsvik R2-0 reactor. Fission fragments were separated with the OSIRIS isotope separator and half-lives were measured with 20 $^3$He neutron counters. ``Mass number 81: The only activity at this mass can be assigned to gallium for the same reasons as in the case of mass 80... Mass number 82: One activity can be found, and again, gallium seems to be the most probable element assignment... Mass number 83: ...We have found only one activity of half-life 0.31$\pm$0.01~s. This activity can be assigned to gallium, as $^{83}$Ge is reported to be a precursor with half-life of $1.9\pm0.4$~s.'' The reported half-lives of 1.23(1)~s ($^{81}$Ga), 0.60(1)~s ($^{82}$Ga), and 0.31(1)~s ($^{83}$Ga) agree with the currently accepted values of 1.217(5)~s, 599(2)~ms, and 308.1(10)~ms, respectively. Rudstam and Lund did not comment on the 2.2~s half-life for $^{81}$Ga \cite{1975Ale01} reported by the OSIRIS collaboration less than three month earlier.

\subsection*{$^{84}$Ga}
The discovery of $^{84}$Ga was reported by Kratz et al.\ in ``Neutron-rich isotopes around the r-process `waiting-point' nuclei $^{79}_{29}$Cu$_{50}$ and $^{80}_{30}$Zn$_{50}$'' in 1991 \cite{1991Kra01}. A $^{238}$UC-graphite target was irradiated with 600 MeV protons from the CERN synchrocyclotron and the fragments were separated and identified with the ISOLDE on-line mass separator. ``During the experiment, three further new isotopes could be identified, i.e. $^{77}$Cu, $^{81}$Zn, and $^{84}$Ga, the latter two lying even `beyond' the r-process path...'' The reported half-life of 85(10)~ms is the currently accepted value.

\subsection*{$^{85,86}$Ga}
Bernas et al.\ observed $^{85}$Ga and $^{86}$Ga for the first time in 1997 as reported in their paper ``Discovery and cross-section measurement of 58 new fission products in projectile-fission of 750$\cdot$A MeV $^{238}$U'' \cite{1997Ber01}. Uranium ions were accelerated to 750 A$\cdot$MeV by the GSI UNILAC/SIS accelerator facility and bombarded a beryllium target. The isotopes produced in the projectile-fission reaction were separated using the fragment separator FRS and the nuclear charge Z for each was determined by the energy loss measurement in an ionization chamber. ``The mass identification was carried out by measuring the time of flight (TOF) and the magnetic rigidity B$\rho$ with an accuracy of 10$^{-4}$.'' 61 and 4 counts of $^{85}$Ga and $^{86}$Ga were observed, respectively.

\subsection*{$^{87}$Ga}
The discovery of $^{87}$Ga was reported in the 2010 article ``Identification of 45 new neutron-rich isotopes produced by in-flight fission of a $^{238}$U Beam at 345 MeV/nucleon,'' by Ohnishi et al.\ \cite{2010Ohn01}. The experiment was performed at the RI Beam Factory at RIKEN, where the new isotopes were created by in-flight fission of a 345 MeV/nucleon $^{238}$U beam on a beryllium target. $^{87}$Ga was separated and identified with the BigRIPS superconducting in-flight separator. The results for the new isotopes discovered in this study were summarized in a table. Ten counts were recorded for $^{87}$Ga.

\section{Discovery of $^{60-90}$Ge}

Thirty-one germanium isotopes from A = 60$-$90 have been discovered so far; these include 5 stable ($^{70}$Ge, $^{72-74}$Ge, and $^{76}$Ge), 11 neutron-deficient and 15 neutron-rich isotopes. According to the HFB-14 model \cite{2007Gor01}, $^{105}$Ge should be the last odd-even particle stable neutron-rich nucleus while the even-even particle stable neutron-rich nuclei should continue through $^{110}$Ge. At the proton dripline two more isotopes ($^{59}$Ge and $^{58}$Ge) should be particle stable and in addition $^{57}$Ge could have a half-life longer than 10$^{-21}$~s \cite{2004Tho01}. Thus, about 21 isotopes have yet to be discovered corresponding to 40\% of all possible germanium isotopes.

\begin{figure}
	\centering
	\includegraphics[scale=.7]{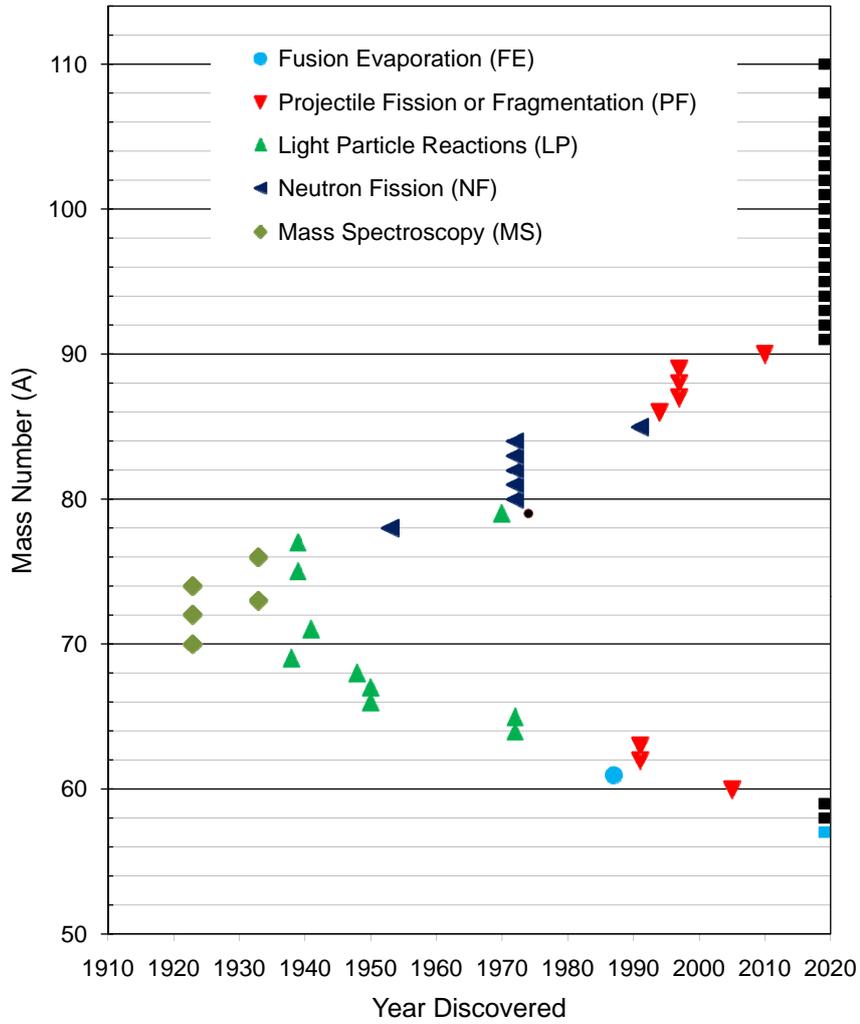}
	\caption{Germanium isotopes as a function of time when they were discovered. The different production methods are indicated. The solid black squares on the right hand side of the plot are isotopes predicted to be bound by the HFB-14 model. On the proton-rich side the light blue square corresponds to an unbound isotope predicted to have a half-life larger than $\sim 10^{-9}$~s. The black solid circle indicates the discovery year of the ground state in the case where the first observation was an isomeric state.}
\label{f:year-germanium}
\end{figure}

Figure \ref{f:year-germanium} summarizes the year of first discovery for all germanium isotopes identified by the method of discovery. The range of isotopes predicted to exist is indicated on the right side of the figure. The radioactive germanium isotopes were produced using fusion-evaporation reactions (FE), light-particle reactions (LP), neutron capture reactions (NC), and projectile fragmentation or projectile fission (PF). The stable isotopes were identified using mass spectroscopy (MS). In the following the discovery of each germanium isotope is discussed in detail and a summary is presented in Table 1.

\subsection*{Stable isotopes $^{70}$Ge, $^{72-74}$Ge, and $^{76}$Ge}
In 1923 Aston reported the discovery of stable $^{70}$Ge, $^{72}$Ge, and $^{74}$Ge in ``The isotopes of germanium'' \cite{1923Ast03}. A pure germanium oxide sample was transformed into a fluorine compound and used in the Cavendish mass spectrograph. ``The effects are somewhat feeble, but satisfactory evidence of the three isotopes has been obtained. Their mass-lines are at 70, 72, 74, and appear to be whole numbers though the accuracy of measurements is not so high as usual.''

In the 1933 paper ``The masses of atoms and the structure of atomic nuclei'' Bainbridge identified stable $^{73}$Ge and $^{76}$Ge \cite{1933Bai01}. $^{73}$Ge was observed with a magnetic spectrograph at the Bartol Research Foundation of the Franklin Institute. In a microphotometer record the germanium isotopes 70, 72, 73, 74, and 76 are clearly visible. The previous observation of $^{73}$Ge by Aston \cite{1928Ast02} was not considered as a discovery because Aston also incorrectly reported the existence of $^{71}$Ge, $^{75}$Ge, and $^{77}$Ge. Bainbridge stated in a conference abstract \cite{1933Bai02}: ``Lines of mass 71, 75 and 77 attributed to Ge$^{73}$, Ge$^{75}$, and Ge$^{77}$ by Aston are mainly if not entirely hydrides of Ge$^{70}$, Ge$^{74}$, and Ge$^{76}$.''

\subsection*{$^{60}$Ge}
In the 2005 paper ``First observation of $^{60}$Ge and $^{64}$Se'' Stolz et al.\ identified $^{60}$Ge \cite{2005Sto01}. $^{60}$Ge was produced in the projectile fragmentation reaction of a 140 MeV/nucleon $^{78}$Kr beam on a beryllium target at the Coupled Cyclotron Facility of the National Superconducting Cyclotron Laboratory at Michigan State University. The projectile fragments were identified with the A1900 fragment separator. ``Three events of $^{60}$Ge were unambiguously identified during 60 hours of beam on target with an average primary beam current of 3.6~pnA.''

\subsection*{$^{61}$Ge}
In ``Beta-delayed proton decay of $^{61}$Ge'' Hotchkis et al.\ reported the observation of $^{61}$Ge in 1987 \cite{1987Hot01}. The Berkeley 88-in.\ cyclotron was used to bombard calcium targets with 77$-$120~MeV $^{24}$Mg beams and $^{61}$Ge was produced in the $^{40}$Ca($^{24}$Mg,3n) fusion-evaporation reaction. Beta-delayed protons were measured with a silicon telescope in connection with a pulsed beam technique and a helium-jet transport system. ``A peak is observed at $E_{lab} = 3.10\pm0.07$ MeV, which can be attributed to the decay of $^{61}$Ge. At 78 MeV this peak is also evident, with a yield two-thirds of that at 85 MeV. Its half-life was deduced to be 40$\pm$15~ms using data from both bombardments.'' This half-life is included in the average accepted value of 44(6)~ms.

\subsection*{$^{62,63}$Ge}
Mohar et al.\ first observed $^{62}$Ge and $^{63}$Ge in the 1991 paper ``Identification of new nuclei near the proton-dripline for 31$\leq$Z$\leq$38'' \cite{1991Moh01}. A 65 A$\cdot$MeV $^{78}$Kr beam produced by the Michigan State K1200 cyclotron reacted with an enriched $^{58}$Ni target. $^{62}$Ge and $^{63}$Ge were identified by measuring the rigidity, $\Delta$E, E$_{total}$, and velocity in the A1200 fragment separator. ``Several new isotopes at or near the proton-drip line are indicated in the mass spectra: $^{61}$Ga, $^{62}$Ge, $^{63}$Ge, $^{65}$As, $^{69}$Br, and $^{75}$Sr.''

\subsection*{$^{64,65}$Ge}
$^{64}$Ge and $^{65}$Ge were first reported by Robertson and Austin in their 1972 paper, ``Germanium-64'' \cite{1972Rob01}. Enriched $^{64}$Zn was irradiated with a 50 MeV $^3$He beam from the Michigan State University cyclotron. Gamma-rays were measured with a Ge(Li) detector following chemical separation. ``The strongest lines decay with a short half-life ($30\pm2$ sec) and are attributed to $^{65}$Ge (despite large disagreement with the previous half-life), on the basis of the rapid growth of the $^{65}$Ga daughter and good energy fit with levels observed in $^{64}$Zn($^3$He,d)$^{65}$Ga. Other lines, at 128.2$\pm$0.2, 384.1$\pm$0.3, 427.0$\pm$0.3, 667.1$\pm$0.3, and 774.5$\pm$0.3 decay with a (weighted average) half-life of 62.3$\pm$2.0 sec and are assigned to the decay of the new isotope $^{64}$Ge.'' These half-lives agree with the presently adopted values of 63.7(25)~s and 30.9(5)~s for $^{64}$Ge and $^{65}$Ge, respectively. The previously reported half-life of $^{65}$Ge mentioned in the quote (1.5(2)~min) was evidently incorrect \cite{1958Por01} and an earlier attempt to observe $^{64}$Ge was unsuccessful \cite{1972deJ01}.

\subsection*{$^{66,67}$Ge}
In the 1950 paper ``Spallation products of arsenic with 190~MeV deuterons'' Hopkins identified $^{66}$Ge and $^{67}$Ge \cite{1950Hop01}. A pure $^{75}$As target was bombarded with 190~MeV deuterons from the Berkeley 184-inch cyclotron. X-rays and $\beta$-rays were recorded following chemical separation. ``The use of improved chemical separations and counting techniques has enabled the identification of 38 nuclear species among the elements from chromium through selenium.'' The half-lives of $\sim$150~min ($^{66}$Ge) and 21~min ($^{67}$Ge) were listed in a table and are in agreement with the presently adopted values of 2.26(5)~h and 18.9(3)~m, respectively. The nominal half-lives listed in the table were quoted from the 1948 Table of Isotopes \cite{1948Sea01} which referred to unpublished data by Hopkins. A previously reported half-life of 195~d tentatively assigned to $^{67}$Ge \cite{1938Man01} was evidently incorrect.

\subsection*{$^{68}$Ge}
Hopkins et al.\ reported the observation of $^{68}$Ge in the 1948 paper ``Nuclear reactions of arsenic with 190-Mev deuterons'' \cite{1948Hop01}. A pure $^{75}$As target was bombarded with 190~MeV deuterons from the Berkeley 184-inch cyclotron. The resulting activities were measured with an argon-filled Geiger-M\"uller counter following chemical separation. The observed half-life of 250~d listed in a table agrees with the presently adopted value of 270.95(16)~d. The 195~d half-life from the literature listed in the table were quoted from the 1944 Table of Isotopes \cite{1944Sea01} which had assigned this half-life to $^{69}$Ge based on a measurement by Mann who had tentatively assigned it to $^{67}$Ge \cite{1938Man01}.

\subsection*{$^{69}$Ge}
The first identification of $^{69}$Ge was made by Mann in 1938, titled ``Nuclear transformations produced in zinc by alpha-particle bombardment'' \cite{1938Man01}. A zinc target was irradiated with 17~MeV $\alpha$-particles from the Berkeley cyclotron. Decay curves and absorption spectra were recorded with two electroscopes following chemical separation. ``The 37-hour activity is probably to be identified with that of 26 hours reported by Sagane for Ge$^{69}$. While the possibility of isomorphism cannot be overlooked, the identification of Ge$^{69}$ seems reasonable.'' The reported half-life agrees with the currently accepted value of 39.05(10)~h. No reference to the 26~h half-life reported by Sagane was given in the paper, but it probably refers to reference \cite{1938Sag02}. However, in that paper, Sagane assigned this half-life to $^{71}$Ge, reporting a half-life of 30~min for $^{69}$Ge. In 1941, Seaborg et al.\ reassigned a 195-d half-life originally assigned to $^{67}$Ge \cite{1938Man01} incorrectly to $^{69}$Ge \cite{1941Sea01}.


\subsection*{$^{71}$Ge}
Seaborg et al.\ correctly identified $^{71}$Ge in the 1941 paper, ``Radioactive isotopes of germanium'' \cite{1941Sea01}. Gallium and germanium targets were bombarded with 8 MeV and 16 MeV deuterons, respectively. Decay curves and absorption spectra were measured. ``Bombarding gallium with 8-Mev deuterons we found in the germanium fraction a 40$\pm$2-hour positron-emitter and an 11-day activity. These are almost certainly due to Ge$^{69}$ or Ge$^{7l}$, produced by the d,2n reaction from the only stable gallium isotopes, Ga$^{69}$ and Ga$^{71}$ (d,n reactions would lead to stable germanium isotopes). But since the bombardment of germanium with 16-Mev deuterons also produces a 40$\pm$2-hour and an 11-day germanium period, both periods must be assigned to Ge$^{71}$, formed by the d,p reaction from Ge$^{70}$.'' The 11~d activity agrees with the currently accepted value of 11.43(3)~d. Seaborg et al.\ also extracted an 11~d half-life from a figure in a paper by Mann \cite{1938Man01} who had not mentioned it. The 40(2)~h activity, however,  most likely corresponds to the half-life of $^{69}$Ge. A previously assigned half-life of 26(3)~h \cite{1939Sag01} was evidently incorrect. Also Aston had incorrectly reported $^{75}$Ge to be stable \cite{1928Ast02}.

\subsection*{$^{75,77}$Ge}
Sagane identified $^{75}$Ge and $^{77}$Ge in 1939 as reported in ``Radioactive isotopes of Cu, Zn, Ga and Ge'' \cite{1939Sag01}. Metallic germanium targets were irradiated by slow and fast neutrons produced by deuteron bombardments of lithium and beryllium from the Berkeley cyclotron. $^{75}$Ge and $^{77}$Ge were produced by neutron capture reactions and the resulting activities were measured with a Lauritsen-type quartz fiber electroscope. ``The 81-min.\ period was formed in strong intensity in each bombardment. This isotope emits negative electrons and is sensitive to slow neutrons. Because of the relative abundance of Ge$^{74}$ (37 percent) and Ge$^{76}$ (6.5 percent), the isotope in question is probably Ge$^{75}$... The 8-hr. period was obtained appreciably only in slow neutron bombardments. With fast neutrons only a trace of this period was noticed, indicating very clearly that this period is very sensitive to slow neutrons. There remains only one possibility for this kind of negative electron active period, that is Ge$^{77}$.'' The reported half-lives of 81(3)~min for $^{75}$Ge and 8(1)~h for $^{77}$Ge are close to the currently accepted values of 82.78(4)~min and 11.30(1)~h, respectively. Previously, Aston had incorrectly reported $^{75}$Ge and $^{77}$Ge to be stable \cite{1928Ast02}.

\subsection*{$^{78}$Ge}
$^{78}$Ge was reported in 1953 by Sugarman in ``Genetics of the Ge$^{78}$--As$^{78}$ fission chain'' \cite{1953Sug01}. Uranyl nitrate was irradiated with thermal neutrons at the Los Alamos Homogenous Reactor. Activities were measured with a methane-flow proportional counter following chemical separation. ``The half-life of Ge$^{78}$ thus found in the two experiments is 86.5~min and 85.3~min. The half-life accepted is 86.0$\pm$1.0~min, if account is taken both of the spread of the data and possible systematic errors... The previously reported value for the half-life of Ge$^{78}$ of 2.1~hr (126 min), as determined directly on a sample of germanium in which arsenic activity was growing, is probably in error because of the difficulty in determining the parent half-life in a mixed sample when the parent and daughter have very nearly the  same half-life.'' The reported half-life agrees with the currently accepted value of 88.0(10)~m. The incorrect half-life measurement of 2.1~h had been reported as part of the Plutonium Project \cite{1951Ste01}.

\subsection*{$^{79}$Ge}
The identification of $^{79}$Ge was reported by Karras et al.\ in ``Radioactive nucleides $^{79}$Ge and $^{82}$As'' in 1969 \cite{1970Kar01}. The Arkansas 400~kV Cockcroft-Walton linear accelerator was used to produce 14.7~MeV neutrons by the $^3$H($^2$H,n)$^4$He reaction. The neutrons irradiated enriched $^{82}$Se samples and $\gamma$- and $\beta$-rays were measured with Ge(Li) and NaI(TI) detectors, respectively. ``The 42~sec activity with $\beta$-rays of 4.3 and 4.0 and the coincident $\gamma$-ray of 230 keV can be assigned to $^{79}$Ge based on the following: (i) the $\beta$-decay Q-value of $4.3\pm0.2$ MeV is in agreement with the estimated value of 4.0 MeV for $^{79}$Ge. $^{82}$As is estimated to have a Q$_\beta\approx$ 7.4 MeV; (ii) the 42~sec activity was found to be considerably enhanced when enriched $^{82}$Se was bombarded and analyzed; and (iii) the 42 sec 230 keV $\gamma$-ray was observed to follow the Ge fraction separated from irradiated natural Se.'' The reported half-life of 42(2)~s agrees with the currently adopted value of 39.0(10)~s for an isomer. The ground-state of $^{29}$Ge was first observed in 1974 \cite{1974Gra01}.

\subsection*{$^{80-84}$Ge}
In the 1972 paper ``Identification of new germanium isotopes in fission: Decay properties and nuclear charge distribution in the A = 78 to 84 mass region'' del Marmol and Fettweis identified $^{80}$Ge, $^{81}$Ge, $^{82}$Ge, $^{83}$Ge, and $^{84}$Ge \cite{1972del01}. A uranyl nitrate solution of $^{235}$U was irradiated with neutrons from the Mol BR1 graphite reactor. Gamma-ray spectra were recorded with a Ge(Li) detector following chemical separation. ``By combining the three methods (the 666.2 keV growth and decay measurement, the `milking' and decay of the 265.6 keV $\gamma$-ray) an average half-life of 24.5$\pm$1.0~s was chosen for $^{80}$Ge, which confirms the assignment of the Osiris group... From the results of both $\beta$- and $\gamma$-ray measurements an average half-life of 10.1$\pm$0.8~s was chosen for $^{81}$Ge... By applying the `milking' method to this same $\gamma$-ray, a half-life of 4.60$\pm$0.35~s was found for $^{82}$Ge as shown in [the figure]... A half-life of 1.9$\pm$0.4~s was determined for $^{83}$Ge by applying the `milking' method to the 356 keV $\gamma$-ray from 22.6~min $^{83}$Se after complete decay of its predecessors and of 70~s $^{83m}$Se... In the present case the `milking' method was applied to the 881.6 keV $\gamma$-ray from $^{84}$Br to obtain the half-life of $^{84}$Ge; even with a counting time of 1~h, the statistics of this 50\% intensity transition were very low.'' The reported half-lives of 24.5(10)~s ($^{80}$Ge), 10.1(8)~s ($^{81}$Ge), 4.60(35)~s ($^{82}$Ge), 1.9(4)~s ($^{83}$Ge), and 1.2(3)~s ($^{84}$Ge) are consistent with the presently adopted values of 29.5(4)~s, 7.6(6)~s, 4.55(5)~s, 1.85(6)~s and 0.954(14)~s, respectively. The $^{82}$Ge half-life is included in the calculation of the average accepted value. The previous assignment of $^{80}$Ge by the OSIRIS group mentioned in the quote was only published in a conference proceeding \cite{1970Gra01}.

\subsection*{$^{85}$Ge}
In 1991 Omtvedt et al.\ described the observation of $^{85}$Ge in ``Gamma-ray and delayed neutron branching data for the new or little known isotopes $^{84,85}$Ge and $^{84,85}$As'' \cite{1991Omt01}. Fission fragments were measured with the OSIRIS facility in Studsvik. Beta- and gamma-ray spectra were recorded of the mass-separated fragments. ``The branchings of $\gamma$-rays following the decay of $^{84,85}$Ge and $^{84,85}$As have been determined using mass-separated samples. Our results include the first identification of $^{85}$Ge and the first $\gamma$-ray data for $^{84}$Ge.'' The reported half-life of 0.58(5)~s agrees with the currently accepted value of 535(47)~ms.

\subsection*{$^{86}$Ge}
Bernas et al.\ discovered $^{86}$Ge in 1994 as reported in ``Projectile fission at relativistic velocities: A novel and powerful source of neutron-rich isotopes well suited for in-flight isotopic separation'' \cite{1994Ber01}. The isotopes were produced using projectile fission of $^{238}$U at 750 MeV/nucleon on a lead target. ``Forward emitted fragments from $^{80}$Zn up to $^{155}$Ce were analyzed with the Fragment Separator (FRS) and unambiguously identified by their energy-loss and time-of-flight.'' The experiment yielded ten individual counts of $^{86}$Ge.

\subsection*{$^{87-89}$Ge}
Bernas et al.\ observed $^{87}$Ge, $^{88}$Ge, and $^{89}$Ge for the first time in 1997 as reported in their paper ``Discovery and cross-section measurement of 58 new fission products in projectile-fission of 750$\cdot$A MeV $^{238}$U'' \cite{1997Ber01}. Uranium ions were accelerated to 750 A$\cdot$MeV by the GSI UNILAC/SIS accelerator facility and bombarded a beryllium target. The isotopes produced in the projectile-fission reaction were separated using the fragment separator FRS and the nuclear charge Z for each was determined by the energy loss measurement in an ionization chamber. ``The mass identification was carried out by measuring the time of flight (TOF) and the magnetic rigidity B$\rho$ with an accuracy of 10$^{-4}$.''  583, 67 and 11 counts were observed for $^{87}$Ge, $^{88}$Ge, and $^{89}$Ge, respectively.

\subsection*{$^{90}$Ge}
The discovery of $^{90}$Ga was reported in the 2010 article ``Identification of 45 new neutron-rich isotopes produced by in-flight fission of a $^{238}$U Beam at 345 MeV/nucleon,'' by Ohnishi et al.\ \cite{2010Ohn01}. The experiment was performed at the RI Beam Factory at RIKEN, where the new isotopes were created by in-flight fission of a 345 MeV/nucleon $^{238}$U beam on a beryllium target. $^{90}$Ge was separated and identified with the BigRIPS superconducting in-flight separator. The results for the new isotopes discovered in this study were summarized in a table. Three counts were recorded for $^{90}$Ge.

\section{Discovery of $^{150-184}$Lu}

Thirty five lutetium isotopes from A = 150$-$184 have been discovered so far; these include 1 stable ($^{175}$Lu), 25 proton-rich and 9 neutron-rich isotopes. According to the HFB-14 model \cite{2007Gor01}, $^{234}$Lu should be the last odd-odd particle stable neutron-rich nucleus while the even-odd particle stable neutron-rich nuclei should continue at least through $^{241}$Lu. The proton dripline has been crossed with the observation of proton emission of $^{150}$Lu and $^{151}$Lu. However, six more isotopes ($^{144-149}$Lu) could possibly still have half-lives longer than 10$^{-9}$~ns \cite{2004Tho01}. Thus, about 60 isotopes have yet to be discovered corresponding to 63\% of all possible lutetium isotopes.

The naming of the element lutetium has been controversial for a long time \cite{2011wik01}. Urbain and Auer von Welsbach claimed the discovery almost simultaneously in 1907/1908. While Urbain named the new element lut\`ecium \cite{1907Urb01} Auer von Welsbach requested the name cassiopeium \cite{1908Aue01}. The international commission on atomic weights - which Urbain was a member of - recommended the name lutetium in 1909 \cite{1909Cla01}. Auer von Welsbach \cite{1909Aue01} and Urbain \cite{1910Urb01} continued to argue about the naming rights in the subsequent years and the name cassiopeium was used in the german scientific literature for several decades (for example \cite{1946Bot01}). In 1949, at the 15$^{th}$ Conference of the International Union of Chemistry it was decided that the spelling should be lutetium rather than lutecium \cite{1949IUC01}.

\begin{figure}
	\centering
	\includegraphics[scale=.7]{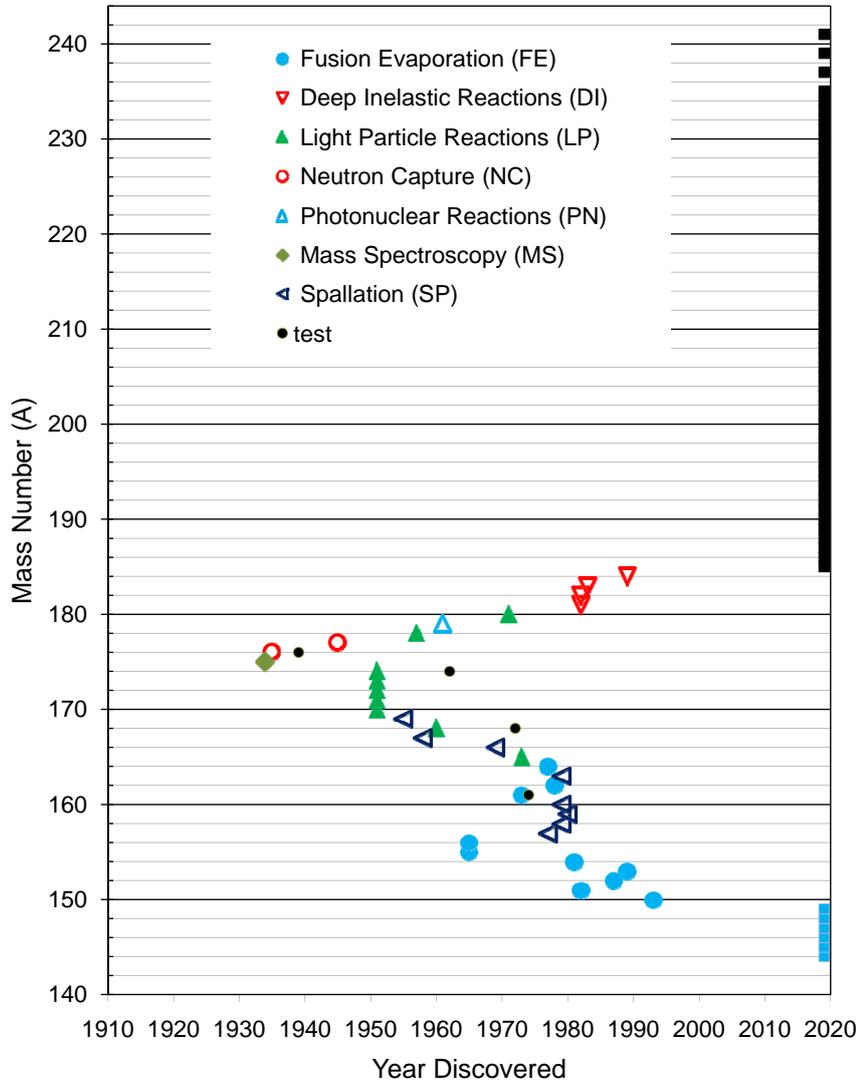}
	\caption{Lutetium isotopes as a function of time when they were discovered. The different production methods are indicated. The solid black squares on the right hand side of the plot are isotopes predicted to be bound by the HFB-14 model. On the proton-rich side the light blue squares correspond to unbound isotopes predicted to have half-lives larger than $\sim 10^{-9}$~s. The black solid circles indicate the discovery year of the ground states in the case where the first observation was an excited state.}
\label{f:year-lutetium}
\end{figure}

Figure \ref{f:year-lutetium} summarizes the year of first discovery for all lutetium isotopes identified by the method of discovery. The range of isotopes predicted to exist is indicated on the right side of the figure. The radioactive lutetium isotopes were produced using fusion-evaporation reactions (FE), photo-nuclear reactions (PN), neutron capture reactions (NC), light-particle reactions (LP), spallation (SP), and deep inelastic (DI). The stable isotope was identified using mass spectroscopy (MS). In the following, the discovery of each lutetium isotope is discussed in detail and a summary is presented in Table 1.

\subsection*{Stable isotope $^{175}$Lu}

Aston discovered $^{175}$Lu in the 1934 paper ``Constitution of dysprosium, holmium, erbium, thulium, ytterbium and lutecium'' \cite{1934Ast04}. Rare earth elements were measured with the Cavendish mass spectrograph: ``Lutecium (71) is simple 175.''

\subsection*{$^{150}$Lu}

$^{150}$Lu was first identified by Sellin et al.\ in their paper ``Proton spectroscopy beyond the drip line near A=150'' in 1993 \cite{1993Sel02}. An isotopically enriched $^{96}$Ru target was bombarded with 300 and 311~MeV $^{58}$Ni beams and $^{150}$Lu was populated in (p3n) fusion evaporation reactions. Recoil products were separated with the Daresbury recoil separator and implanted in a double-sided silicon strip detector where subsequent $\alpha$-particle and proton-decays were measured. ``The assignment of this proton transition to the p3n evaporation channel from $^{154}$Hf$^*$ is also consistent with the observed yield and we therefore identify the 1.26-MeV peak as proton emission from $^{150}$Lu.'' The reported half-life of 35(10)~ms is included in the calculation of the currently accepted value of 46(6)~ms. Earlier reports of a half-life of $\le$10~ms \cite{1984Hof01} and a proton decay energy of 1262.7(36)~keV \cite{1989Hof02} were only reported in conference proceedings.

\subsection*{$^{151}$Lu}

Hoffmann et al.\ discovered $^{151}$Lu as reported in the 1981 paper ``Proton radioactivity of $^{151}$Lu'' \cite{1982Hof01}. The GSI UNILAC accelerator was used to bombard enriched $^{96}$Ru targets with 261$-$302~MeV $^{58}$Ni beams to produce $^{151}$Lu in (p2n) fusion-evaporation reactions. Reaction products were separated with the SHIP velocity separator onto position sensitive surface barrier detectors for energy and time measurements as well as subsequent particle decays. ``We conclude that, barring any effects not foreseen both by the systematics and by theory, the observed Q$_P$ value is highly compatible with emissions from the $^{151}$Lu ground state or an isomeric state very close to it.'' The observed half life of 85(10)~ms is included in the calculation of the presently accepted value of 80.6(20)~ms.

\subsection*{$^{152}$Lu}

The identification of $^{152}$Lu was reported by Toth et al.\ in 1986 in ``Investigation of A = 152 radioactivities with mass-separated sources: Identification of $^{152}$Lu'' \cite{1987Tot01}. An enriched $^{96}$Ru target was bombarded with 354~MeV $^{58}$Ni ions from the Berkeley SuperHILAC and $^{152}$Lu was formed in the (pn) fusion evaporation reaction. The products were separated by the OASIS isotope separator which delivered the isotopes to a $\Delta E-E$ telescope, Ge detectors, and a plastic scintillator with a tape system. ``Three weak $\gamma$ rays, 312.6, 358.7, and 1531.4 keV, known to deexcite $^{152}$Yb levels, were seen in our spectra to decay with a (0.7$\pm$0.1)-s half-life, one that had previously not been observed in the $A=152$ isobaric chain. We therefore attribute them to the $\beta$ decay of the new isotope $^{152}$Lu.'' This half-life corresponds to the currently adopted value.

\subsection*{$^{153}$Lu}

$^{153}$Lu was observed by Nitschke et al.\ in their 1989 paper ``Identification of $^{152}$Lu $\beta$ decay'' \cite{1989Nit01}. A 284~MeV $^{64}$Zn beam from the Berkeley SuperHILAC bombarded an enriched $^{92}$Mo target populating $^{153}$Lu in the (p2n) fusion-evaporation reaction. The online isotope separator OASIS was used to separate the reaction products which were then transported by a cycling tape system to a detection system consisting of a silicon $\Delta E-E$ telescope, a plastic scintillator, a planar hyperpure Ge detector and two n-type Ge detectors. ``Gating on the 566.5-keV $\gamma$ ray revealed coincident Yb K x rays indicating that we had observed the $\beta$ decay of $^{153}$Lu.'' The reported half-life of 0.9(2)~s corresponds to the presently adopted value. In an earlier report the existence of $^{153}$Lu was inferred from $\alpha$-correlation measurements: ``Further, correlations were measured between the $\alpha$ lines of $^{157}$Ta$ - ^{153}$Tm and $^{156}$Hf$ - ^{152}$Er that prove a $\beta$-decay of the new isotopes $^{153}$Lu, $^{152}$Yb, and $^{152}$Tm.'' \cite{1981Hof01}. However, no properties of $^{153}$Lu or its decay were measured.

\subsection*{$^{154}$Lu}

In the 1981 paper ``New neutron deficient isotopes in the range of elements Tm to Pt'' Hofmann et al.\ reported the first observation of $^{154}$Lu \cite{1981Hof01}. Neutron deficient isotopes of elements from molybdenum to tin and vanadium to nickel targets were bombarded with $^{58}$Ni and $^{107}$Ag at the GSI linear accelerator UNILAC. Reaction products were separated by the SHIP velocity filter and implanted into silicon detectors. ``Here, the intensity of the $^{154}$Yb $\alpha$ line first increases corresponding to its own half-life of 410~ms followed by a decrease with T$_{1/2}$~=~960~ms corresponding to the half-life of the $\beta$ emitter $^{154}$Lu.'' The reported half life of 960(100)~ms is included in the calculation of the currently accepted value of 1.12(8)~s.

\subsection*{$^{155,156}$Lu}

Macfarlane discovered $^{155}$Lu and $^{156}$Lu in 1963 as reported in ``Alpha-decay properties of some lutetium and hafnium isotopes near the 82-neutron closed shell'' \cite{1965Mac01}. An enriched $^{144}$Sm target was bombarded with $^{19}$F at the Berkeley heavy-ion linear accelerator and $^{155}$Lu and $^{156}$Lu were produced in (8n) and (7n) fusion evaporation reactions, respectively. Excitation functions and $\alpha$-particles spectra were measured. ``However, the excitation function for the 5.63~MeV alpha group which we suspected to be due to Lu$^{155}$ peaks $\approx$6~MeV lower than this, at a value half-way between the (H.I.,8n) and (H.I.,7n) energies. We did observe an activity, however, which was assigned to the (F$^{19}$,7n) reaction so that the only plausible mass assignment for this activity is Lu$^{155}$. \dots\ The peak of the excitation function falls at an excitation energy of 103~MeV which is in good agreement with the values previously observed for (H.I.,7n) reactions. (See the above discussion of the Lu$^{155}$ results.) The mass assignment of this activity must, therefore, be Lu$^{156}$.'' The measured half life of 0.07(2)~s for $^{155}$Lu is used for the calculation of the presently adopted value of 68(1)~ms. Half lives of $\approx$0.5~s and 0.23(3)~s were reported for the ground state and a high spin state of $^{156}$Lu, which agree with the accepted values of 494(12)~ms and 198(2)~ms, respectively.

\subsection*{$^{157}$Lu}

The observation of $^{157}$Lu was reported by Hagberg et al.\ in the 1977 paper ``Alpha decay of neutron-deficient ytterbium isotopes and their daughters'' \cite{1977Hag02}. The CERN synchro-cyclotron was used to bombard tantalum with 600~MeV protons. $^{157}$Lu was separated with the ISOLDE on-line mass separator facility and $\alpha$ particles were measured with two silicon surface-barrier detectors. ``At this mass number, we have also observed two short-lived activities with $\alpha$-energies 4.98$\pm$0.02~MeV and 5.11$\pm$0.02~MeV. The latter corresponds well to the known $\alpha$-decay energy of $^{153}$Tm and thus we assign the 4.98~MeV activity to its $\alpha$-decay parent $^{157}$Lu.'' Hagberg et al. mention that their result differs significantly from a previous measurement referring to a conference abstract \cite{1972Gau02}.

\subsection*{$^{158}$Lu}

Alkhazov et al.\ identified $^{158}$Lu in the 1979 paper ``New neutron deficient lutetium isotopes'' \cite{1979Alk01}. Tungsten and tantalum targets were bombarded with 1~GeV protons from the Leningrad synchrocyclotron and $^{158}$Lu was produced in spallation reactions. It was separated with the IRIS mass separator and subsequent decays were measured with a surface-barrier detector as well as X- and $\gamma$-ray detectors. ``On the basis of the parent-daughter relationship the E$_\alpha$ = 4.665~MeV line T$_{1/2}$ = 10.4$\pm$1.0~s has been assigned to the new $^{158}$Lu.'' This half-life is included in the calculation of the presently adopted value of 10.6(3)~s.

\subsection*{$^{159}$Lu}

$^{159}$Lu was observed by Alkhazov et al.\ as described in ``New isotope $^{159}$Lu and decay of $^{158}$Lu, $^{159,158}$Yb isotopes'' \cite{1980Alk01}. Tungsten targets were bombarded with 1~GeV protons from the Leningrad synchrocyclotron and $^{159}$Lu was produced in spallation reactions. It was separated with the IRIS mass separator and subsequent decays were measured with a surface-barrier and Ge(Li) detectors. ``The half-life of the new isotope $^{159}$Lu was determined from X-ray, $\gamma$-ray and $\alpha$ particle measurement. [The figure] shows the Roentgen spectrum obtained at this isobar and the decay data for K$_{\alpha_1}$Yb. To confirm our earlier preliminary assignment of 4.43~MeV $\alpha$ line ( T$_{1/2}\approx$ 20 s) to $^{159}$Lu we remeasured the $\alpha$ spectrum for this mass with higher statistics. From the present experiment we obtained E$_\alpha$ = 4.420$\pm$0.010~MeV for the energy of this new $\alpha$ line. The measured half-life T$_{1/2}$ = 12.0$\pm$1.5~s is in good agreement with the value obtained by X-ray and y-ray spectroscopy.'' This half-life is included in the calculation of the presently accepted value of 12.1(10)~s. The preliminary assignment was published a year earlier \cite{1979Alk01}.

\subsection*{$^{160}$Lu}

Alkhazov et al.\ identified $^{160}$Lu in the 1979 paper ``New neutron deficient lutetium isotopes'' \cite{1979Alk01}. Tungsten and tantalum targets were bombarded with 1~GeV protons from the Leningrad synchrocyclotron and $^{160}$Lu was produced in spallation reactions. It was separated with the IRIS mass separator and subsequent decays were measured with a surface-barrier detector as well as X- and $\gamma$-ray detectors. ``Isotopes $^{158,160,161,163}$Lu have been identified for the first time... The identification of the new isotopes is based on the analysis of the characteristic K$_\alpha$ and K$_\beta$ lines in the X-ray spectra, and the genetic relationship to the decay of the daughter well known nuclei, in addition to the unambiguous mass determination after mass separations.'' The measured half-life (34.5(15)~s) listed only in a table is included in the calculation of the presently adopted value of 36.1(3)~s.

\subsection*{$^{161}$Lu}

In 1973 the identification of $^{161}$Lu was reported in ``A 7.3~ms isomer of $^{161}$Lu'' by Anholt et al.\ \cite{1973Anh01}. An enriched $^{148}$Sm target was bombarded with 110$-$150~MeV $^{19}$F ions from the Yale heavy ion accelerator and $^{161}$Lu was formed in the (6n) fusion evaporation reaction. Decay curves and $\gamma$-ray spectra of the subsequent activities were measured with a Ge(Li) detector. ``The mass number of the decay chain to which this activity belongs has been established as 161 by observing the excitation function for this line and comparing it to those of $\gamma$-ray peaks associated with radioactive decay products of $^{160}$Lu, $^{161}$Lu and $^{162}$Lu'' The reported half-life of 7.3(4)~ms corresponds to an isomeric state and is the currently adopted value. Six years later Alkhazov et al.\ claimed the discovery of $^{161}$Lu by reporting the measurement of the ground state \cite{1979Alk01}.

\subsection*{$^{162}$Lu}

In ``Decay of $^{162,164,165}$Lu isotopes'' Burman et al.\ reported the discovery of $^{162}$Lu in 1978 \cite{1978Bur01}. A $\sim$105~MeV $^{16}$O beam from the Yale heavy ion accelerator bombarded enriched $^{151}$Eu$_2$O$_3$ powder to produce $^{162}$Lu in the reaction $^{151}$Eu($^{16}$O,5n). Decay curves, $\gamma$-ray singles spectra and $\gamma-\gamma$ coincidence spectra were recorded using multiple Ge(Li) detectors. ``Only two $\gamma$ rays at 167.0 and 320.3 keV have been observed in the decay of $^{162}$Lu to the levels of $^{162}$Yb. The half-life of $^{162}$Lu (1.40$\pm$0.15~min) has been verified in the present work. No gamma rays have been reported earlier from the decay of $^{162}$Lu isotope.'' The currently accepted half-life of 1.37(2)~min was calculated by including this reported half-life. Previous measurements were only published in an internal report \cite{1968Nei01} and a conference proceeding \cite{1976Ern01}.

\subsection*{$^{163}$Lu}

Alkhazov et al.\ identified $^{163}$Lu in the 1979 paper ``New neutron deficient lutetium isotopes'' \cite{1979Alk01}. Tungsten and tantalum targets were bombarded with 1~GeV protons from the Leningrad synchrocyclotron and $^{163}$Lu was produced in spallation reactions. It was separated with the IRIS mass separator and subsequent decays were measured with a surface-barrier detector as well as X- and $\gamma$-ray detectors. ``Isotopes $^{158,160,161,163}$Lu have been identified for the first time... The identification of the new isotopes is based on the analysis of the characteristic K$_\alpha$ and K$_\beta$ lines in the X-ray spectra, and the genetic relationship to the decay of the daughter well known nuclei, in addition to the unambiguous mass determination after mass separations.'' The measured half-life (246(12)~s) listed only in a table is included in the calculation of the presently adopted value of 3.97(13)~min.

\subsection*{$^{164}$Lu}

$^{164}$Lu was observed by Hunter et al.\ in 1977 as described in ``Levels in $^{164}$Yb from $^{164}$Lu decay'' \cite{1977Hun01}. A 79~MeV $^{14}$N beam from the Oak Ridge isochronous cyclotron irradiated gadolinium oxide enriched in $^{155}$Gd. Two Ge(Li) detectors were used to record single and coincidence $\gamma$-ray spectra. ``The best value of the half-life of the dominant $^{164}$Lu decay appears to be 3.17$\pm$0.004~min, obtained from the 123.8-keV peak.'' The reported half-life is included in the calculations of the currently accepted half-life of 3.14(3)~m. Hunter et al.\ mentioned a previously reported 3.1~min half-life which was only published as an internal report \cite{1968Nei01}.

\subsection*{$^{165}$Lu}

The first identification of $^{165}$Lu was reported in ``Gamma rays from 11.8~min $^{165}$Lu, a new isotope'' by Meijer et al.\ in 1973 \cite{1973Mei01}. Thulium foils were bombarded with 50, 60, 70 and 80~MeV $^3$He beams from the Amsterdam synchrocyclotron. A Ge(Li) anti-compton spectrometer and a Ge(Li) surface barrier detector were used to measure the $\gamma$-ray spectra. ``In experiments with 70 or 80~MeV bombarding energy we observed several strong lines decaying with 11.9$\pm$0.5~min. The highest production of this activity was at 80~MeV irradiation energy, corresponding to the maximum of the ($^3$He,7n) reaction. Further evidence for the assignment to $^{165}$Lu is provided by the fact that we observed the strongest lines in the $^{165}$Yb daughter decay growing in with about 9 minutes and decaying with 11.8 minutes.'' This half-life agrees with the currently accepted value of 10.74(10)~min.

\subsection*{$^{166}$Lu}

$^{166}$Lu was observed by Arlt et al.\ as reported in the 1970 paper ``The new neutron-deficient isotopes $^{169}$Hf, $^{167}$Hf, $^{166}$Hf, and $^{166}$Lu and the decay scheme of $^{169}$Hf'' \cite{1969Arl04}. Protons were accelerated to 660~MeV by the Dubna synchrocyclotron and bombarded Ta$_2$O$_5$ targets to form hafnium isotopes in the Ta(p,2pxn) reaction which subsequently decayed to lutetium isotopes. Gamma-ray spectra were measured with NaI(Tl) and Ge(Li) detectors in singles and coincidences following chemical separation. ``The 102, 228, and 338~keV $\gamma$ rays decayed with a 3.3$\pm$0.2~min half-life. This activity is derived from the activity decaying with T=5.8$\pm$0.2~min, i.e.\ from $^{166}$Hf, so we may suppose that it arises from the decay of $^{166}$Lu.'' This half-life is close to the currently accepted value of 2.65(10)~min.

\subsection*{$^{167}$Lu}

Aron et al.\ discovered $^{167}$Lu in ``New neutron-deficient rare earth isotopes. Lutetium isotope with mass number 167'' in 1958 \cite{1958Aro01}. Tantalum was bombarded with 660~MeV protons from the Dubna synchrocyclotron. A scintillation $\gamma$-spectrometer was used to measure $\gamma$-ray spectra following chemical separation. ``In the spectrum of the thulium fraction separated simultaneously with the ytterbium from the lutetium fraction we observed the $\gamma$-lines characteristic of Tm$^{167}$. The intensity of the bright 207~kev $\gamma$-line fell off with a period of $\approx$10~days (the tabular value for the period of Tm$^{167}$ is 9.6 days). Thus we have unquestionable evidence of the existence of a hitherto unknown lutetium isotope, namely Lu$^{167}$.'' The reported half-life of 55(3)~min agrees with the presently adopted value of 51.5(10)~min. In the following year a 54~min half-life was reported independently \cite{1959Har01}.

\subsection*{$^{168}$Lu}

$^{168}$Lu was reported in ``Radioactive decay of Lu$^{168}$'' by Wilson and Pool in 1959 \cite{1960Wil01}. Ytterbium oxide targets enriched in $^{168}$Yb were bombarded with 6~MeV protons and $^{168}$Lu was produced in the (p,n) charge exchange reaction. Decay curves, $\gamma$-ray and K~X-ray spectra were measured. ``The initially resulting activity is assigned to Lu$^{168}$ by the identification of the ytterbium K~x~ray and by comparison with the activities produced by similar proton irradiations of each of the other enriched isotopes of ytterbium.'' The reported half-life of 7.1(2)~min is included in the calculation of the currently accepted value of 6.7(4)~min for an isomeric state. The ground-state of $^{168}$Lu was first observed in 1972 \cite{1972Cha02}.

\subsection*{$^{169}$Lu}

Nervik and Seaborg reported the discovery of $^{169}$Lu in 1954 in ``Tantalum spallation and fission induced by 340-MeV protons'' \cite{1955Ner01}. Tantalum metal was bombarded with 340-MeV protons from the Berkeley 184-inch cyclotron. Decay curves were measured with a Geiger-M\"uller counter following chemical separation. ``It seems more probable that 32-day Yb$^{169}$ is growing into the lutetium fraction as a daughter activity of Lu$^{169}$. From the relative magnitude of the 32-day and 1.7-day activities, the 1.7-day activity could be the Lu$^{169}$ parent.'' This half-life is close to the accepted value of 34.06(5)~h.

\subsection*{$^{170-174}$Lu}

In ``Radioactive isotopes of lutetium and hafnium'' Wilkinson and Hicks described the identification of $^{170}$Lu, $^{171}$Lu, $^{172}$Lu, $^{173}$Lu, and $^{174}$Lu in 1950 \cite{1951Wil01}. Targets of rare earth elements were irradiated with various light particles produced with the Berkeley 60-in.\ cyclotron and the linear accelerator. $^{170}$Lu, $^{171}$Lu, and $^{172}$Lu were primarily produced by bombarding thulium targets with 15$-$38~MeV $\alpha$ particles, $^{173}$Lu with 30$-$40~MeV protons on lutetium, and $^{174}$Lu with 10~MeV protons, 19~MeV deuterons and fast neutrons on lutetium. Decay curves, absorption curves, and electron spectra were measured following chemical separation. ``1.7$\pm$0.1-day Lu$^{170}$ --- This activity was observed in bombardments of thulium with alpha-particles of energy greater than about 30~Mev. It was also found in the decay of a short-lived hafnium parent (112-min Hf$^{170}$). Allocation to mass 170 thus seems fairly certain... 8.5$\pm$0.2-day Lu$^{171}$ --- Growth of the 1.7 day-Lu$^{170}$ together with growth of the 8.5-day species was observed only in hafnium activities produced by 60 to 75-Mev protons on lutetium and hence, in view of the production of both isotopes in alpha particle bombardments of thulium, allocation of the 8.5-day activity is made to mass 171... 6.70$\pm$0.05 day Lu$^{172}$ --- The hafnium activity of about five years half-life allocated to mass 172 has been found to have a lutetium daughter, the decay of which has been followed through over ten half-lives. The radiation characteristics agree well with those obtained for the activity produced in low energy alpha-particle bombardments of thulium, and it is fairly certain that the activities are due to the same isotope... $\sim$500-day Lu$^{173}$ --- The quantity of residual lutetium activity formed in decay of a sample of pure 23.6-hour Hf$^{173}$ made by the (p,3n) reaction in the bombardment of lutetium with 25-Mev protons in the linear accelerator, together with the estimated counting efficiencies, gives a half-life of 400 to 600 days for this isotope. The direct decay has been followed only through two years as yet, yielding a value $\sim$500 days... 165$\pm$5-day Lu$^{174}$ --- This activity was found in the lutetium fraction from bombardments of lutetium with fast neutrons, 10-Mev protons, and 19-Mev deuterons; it has also been found together with the well known 6.7-day Lu$^{177}$ in the bombardment of hafnium with 19-Mev deuterons. Allocation to mass 174 can be made with certainty on the basis of formation by (n,2n), (p,pn), and (d,p2n) reactions in lutetium and the (d,$\alpha$) reaction in hafnium.'' These half-lives are in reasonable agreements with the currently adopted values of 2.012(20)~d ($^{170}$Lu), 8.24(3)~d ($^{171}$Lu), 6.70(3)~d ($^{172}$Lu), 1.37(1)~y ($^{173}$Lu), and 142(2)~d ($^{174}$Lu). The value for $^{174}$Lu corresponds to an isomer. The ground-state of $^{174}$Lu was first observed in 1962 \cite{1962Bon01}.



\subsection*{$^{176}$Lu}

$^{176}$Lu was identified by Marsh and Sugden and published in the 1935 paper ``Artificial radioactivity of the rare earth elements'' \cite{1935Mar01}. Ytterbium and lutetium oxides were irradiated with neutrons from a 400 mCi radon source in contact with powdered beryllium. ``We have now examined a specimen of ytterbia separated from lutecia and other earths by the electrolytic method as insoluble YbSO$_4$. It gives a very feeble activity  which is indistinguishable in period from that of lutecium and is probably due to residual traces of that element. The residual earths after the separation of ytterbium consisted chiefly of lutecia and gave a strong activity identical in period with that found for Prof.\ Urbain's specimen of lutecia.'' The reported half-life of 4.0(1)~h is close to the accepted half-life of 3.664(19)~h. This corresponds to an isomeric state. $^{176}$Lu can be considered stable with a half-life of 38.5(7)~Gy. The ground-state of $^{176}$Lu was first observed in 1939 \cite{1939Lib01}.

\subsection*{$^{177}$Lu}

The identification of $^{177}$Lu was reported by Atterling et al.\ in 1944 in ``Neutron-induced radioactivity in lutecium and ytterbium'' \cite{1945Att01}. Lu$_2$O$_3$ samples were bombarded with fast and slow neutrons produced by bombarding LiOH with 6~MeV deuterons and beryllium with 6.5 MeV deuterons from the Stockholm cyclotron, respectively. The resulting activities were measured with a Wulf string electrometer and a Geiger-M\"uller counter. ``'We find two periods in lutecium, 3.67~h. and 6.6 d. With slow neutrons both periods are activated strongly, the 6.6 d. period being the stronger as regards the saturation activity. With fast neutrons, however, the 3.67~h. period is much more strongly activated than the long-lived activity... As will be seen from the isotope-diagram in [the figure], the above mentioned variation in the activities of the two periods caused by the different methods of activation very strongly indicates that the 6.6~d.\ period should be assigned to Lu$^{177}$.'' The reported half-life of 6.6(1)~d agrees with the currently accepted value of 6.647(4)~d. Previously a 5~d activity was measured without a specific mass assignment \cite{1935Hev01} and 6~d \cite{1938Poo02} and 6$-$7~d \cite{1936Hev01} were assigned to $^{176}$Lu. Flammersfeld and Bothe assigned a 6.6~d activity to $^{176}$Lu and a 3.4~h activity to $^{177}$Lu \cite{1943Fla01}.

\subsection*{$^{178}$Lu}
In 1957, Stribel described the discovery of $^{178}$Lu in ``Massenzuordnung und $\gamma$-Spektrum des 22 min-Lutetium'' \cite{1957Str01}. A metallic tantalum target was irradiated with fast neutrons produced by bombarding deuterons on lithium and $^{178}$Lu was produced in (n,$\alpha$) reactions. The resulting activity was measured with a NaI scintillation spectrometer. ``Bei Diskriminierung auf $\gamma$-Energien \"uber 250 keV fanden wir einen zeitlichen Abfall von etwa 20 min Halbwertszeit. Eine chemische Abtrennung wurde nicht durchgef\"uhrt. Da jedoch andere Aktivit\"aten \"ahnlicher Periode mit schnellen Neutronen nicht entstehen k\"onnen, d\"urfte diese $\gamma$-Aktivit\"at mit dem 22 min. Lutetium identisch sein, dem danach die Massenzahl 178 zuzuordnen w\"are.'' [When gating on $\gamma$-ray energies larger than 250 keV, we found a decay with a half-life of about 20~min. No chemical separation was performed. However, because no other activities with similar lifetime can be produced with fast neutrons this $\gamma$-activity has to be identical with the 22-min lutetium and assigned to mass 178.] Previously, half-lives of 22~min and 8~h were assigned to either $^{178}$Lu or $^{179}$Lu \cite{1950But01}.

\subsection*{$^{179}$Lu}

Kuroyanagi et al. observed $^{179}$Lu in 1961 as reported in ``New activities in rare earth region produced by the ($\gamma$,p) reactions'' \cite{1961Kur01}. Hafnium oxide powder was irradiated with $\gamma$-rays at the Tohoku 25 MeV betatron. Decay curves were measured with a $\beta$-ray analyser or an end-window G-M counter and $\beta$-ray spectra were recorded with a plastic scintillator. ``Several new activities in the rare earth region were identified and also the decay characteristics of some previously reported activities in this region were studied more in detail. They were produced by the ($\gamma$,p) reactions and measurement was made with the aid of the scintillation spectrometers. Results were as follows: ... Lu$^{179}$: 7.5$\pm$0.5~h (half-lives), 1.35$\pm$0.05~meV (beta rays), 90, 215~keV (gamma rays).'' This half-life is in reasonable agreement with the presently adopted value of 4.59(6)~h. A year later Stensland and Voigt confirmed the results except for the 90~keV $\gamma$-ray and attributed the longer half-life to possible impurities \cite{1963Ste01}. Previously, half-lives of 22~min and 8~h were assigned to either $^{178}$Lu or $^{179}$Lu \cite{1950But01} and a $\sim$5~h half-life was reported without a mass assignment \cite{1951But01}.

\subsection*{$^{180}$Lu}

The identification of $^{180}$Lu was described by Gujrathi and D'Auria in 1970 in ``The decay of $^{180}$Lu to the levels of $^{180}$Hf'' \cite{1971Guj01}. Natural hafnium metal and hafnium oxide enriched in $^{180}$Hf were irradiated with 14.8~MeV neutrons from a Texas Nuclear Corp. Model 9400 neutron generator. Subsequent radiation was measured with a $\gamma$-ray spectrometer, an X-ray spectrometer and an anthracene crystal. ``The remaining resolved photopeaks are found to decay with a composite half-life of 5.3$\pm$0.3~min and are assigned to the decay of a beta unstable level in $^{180}$Lu.'' This half-life is included in the calculation of the currently accepted half-life of 5.7(1)~min. Previously a 4.5(1)~min half-life was assigned to either  $^{178m}$Lu or $^{180}$Lu \cite{1960Ate01} and a 2.5~min half-life \cite{1961Tak01} was evidently incorrect. It is interesting to note that less than three month later Swindle et al.\ reported the discovery of the ``New isotope $^{180}$Lu'' \cite{1971Swi01} quoting the uncertain mass assignment of Aten and Funke-Klopper \cite{1960Ate01}, but then retracted the discovery claim \cite{1971Swi02} stating that it had been brought their attention that Aten and Funke-Klopper had discovered $^{180}$Lu quoting the same paper \cite{1960Ate01}.

\subsection*{$^{181,182}$Lu}

$^{181}$Lu and $^{182}$Lu were discovered by Kirchner et al.\ in 1981 and reported in ``New neutron-rich $^{179}$Yb and $^{181,182}$Lu isotopes produced in reactions of 9~MeV/u $^{136}$Xe ions on tantalum and tungsten targets'' \cite{1982Kir01}. A 9~MeV/u $^{146}$Xe beam from the GSI UNILAC accelerator bombarded tungsten and tantalum targets. $^{181}$Lu and $^{182}$Lu were identified with an online-mass separator and $\beta$-, $\gamma$-, and X-ray decay spectroscopy. ``The half-life of 3.5$\pm$0.3~min for $^{181}$Lu resulted from the analysis of the decay curves of $\beta$-rays, hafnium K X-rays and the $\gamma$-transitions with energies of 205.9 and 652.4~keV... The $^{182}$Lu half-life of 2.0$\pm$0.2~min was obtained from the analysis of the decay-curves of $\beta$-rays, hafnium K X-rays and the $\gamma$-transitions with energies of 97.8 and 720.8~keV.'' These half-lives correspond to the currently accepted values.

\subsection*{$^{183}$Lu}

Rykaczewski et al.\ reported the discovery of $^{183}$Lu in the 1982 paper ``The new neutron-rich isotope $^{183}$Lu'' \cite{1983Ryk01}. The GSI UNILAC accelerator was used to bombard a tungsten/tantalum target with a 11.7~MeV/u $^{136}$Xe beam. A plastic scintillator and two Ge(Li) detectors were used to measure $\beta$ and $\gamma$ spectra, respectively, following on-line mass separation. ``The observed coincidence of hafnium K X-rays with the 168, 249, and 1057~keV $\gamma$-lines allows, together with $\beta-\gamma$ coincidence relationships, to assign these $\gamma$-lines to the decay of $^{183}$Lu.'' The reported half-life of 58(4)~s is the currently accepted half-life.

\subsection*{$^{184}$Lu}

$^{184}$Lu was identified by Rykaczewski et al.\ as reported in ``Investigation of neutron-rich rare-earth nuclei including the new isotopes  $^{177}$Tm and $^{184}$Lu'' in 1989 \cite{1989Ryk01}. A stack of tungsten and tantalum foils were bombarded with 9-15~MeV/u $^{136}$Xe, $^{186}$W, and $^{238}$U beams from the GSI UNILAC accelerator. Plastic scintillators and Ge(Li) detectors were used to measure $\beta$- and $\gamma$-ray spectra, respectively following on-line mass separation. ``No coincidences of 107~keV $\gamma$-rays with $\gamma$-lines other than 368 and 243~keV were found. Thereby, a direct feeding of the 2$^+$ level in $\beta$-decay of a low-spin ground or isomeric state of $^{184}$Lu is very likely. This was confirmed in an additional measurement with a 0.7~mm thick aluminium foil in front of the $\beta$-detector to absorb $\beta$-rays with energies $\le$0.45 MeV. With these high-energy $\beta$-rays, 107~keV $\gamma$-rays and Hf K X-rays were measured in coincidence, which represents clear evidence for $\beta$-decay of a low-spin $^{184}$Lu state to the 107~keV, 2$^+$ level in $^{184}$Hf.'' The reported half-life of $\sim$18~s agrees with the presently adopted value of 19(2)~s.

\section{Discovery of $^{154-189}$Hf}

Thirty six hafnium isotopes from A = 154$-$189 have been discovered so far; these include 6 stable ($^{174}$Hf and $^{176-180}$Hf), 21 proton-rich and 9 neutron-rich isotopes. According to the HFB-14 model \cite{2007Gor01}, $^{235}$Hf should be the last odd-even particle stable neutron-rich nucleus while the even-even particle stable neutron-rich nuclei should continue at least through $^{240}$Hf. The discovery of $^{153}$Hf had been reported in a conference proceeding \cite{2000Sou01} but never in a refereed publication. At the proton dripline at least four more particle stable hafnium isotopes are predicted ($^{150-153}$Hf) and in addition seven more isotopes ($^{143-149}$Hf) could possibly still have half-lives longer than 10$^{-9}$~s \cite{2004Tho01}. Thus, about 60 isotopes have yet to be discovered corresponding to 63\% of all possible hafnium isotopes.

\begin{figure}
	\centering
	\includegraphics[scale=.7]{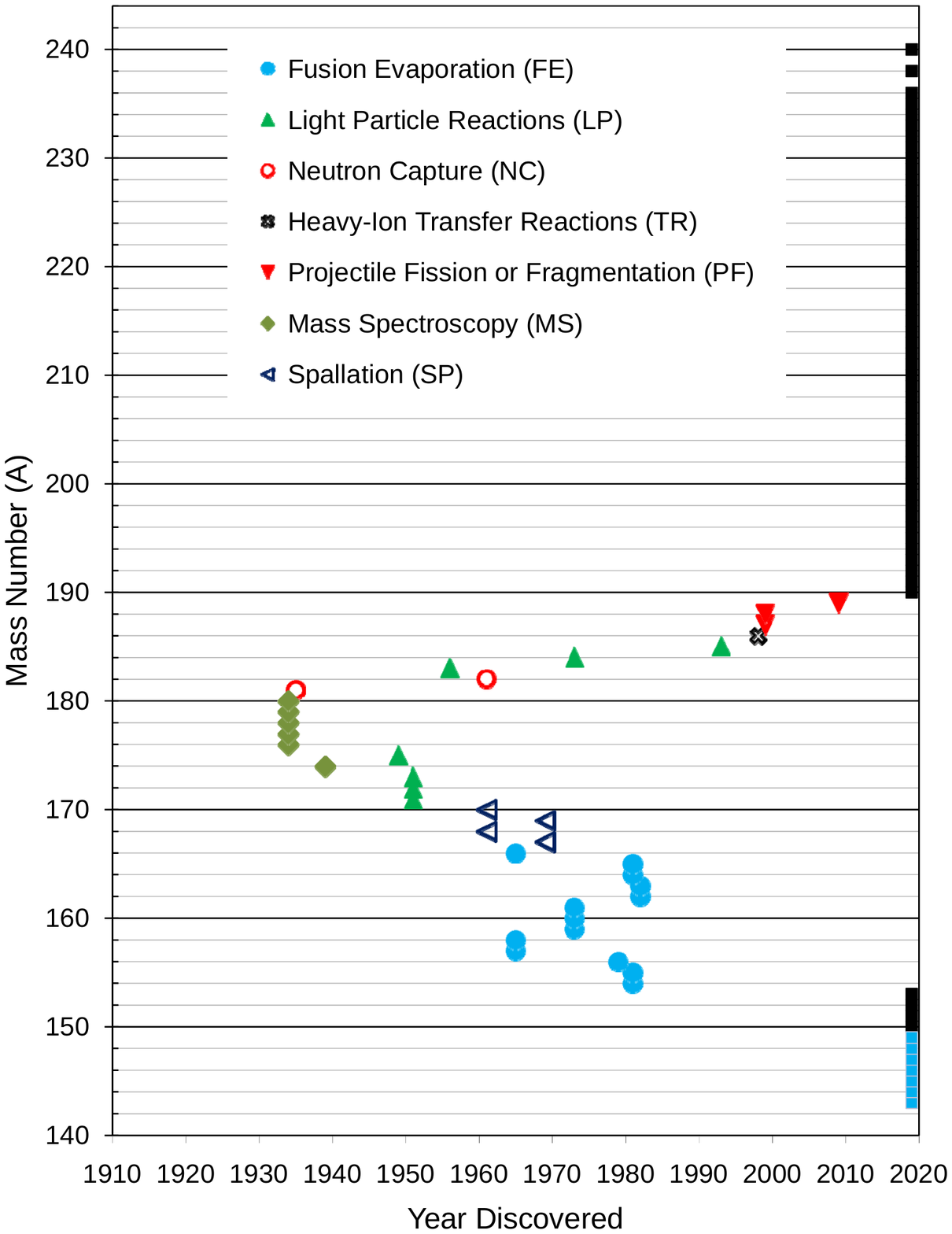}
	\caption{Hafnium isotopes as a function of time when they were discovered. The different production methods are indicated. The solid black squares on the right hand side of the plot are isotopes predicted to be bound by the HFB-14 model. On the proton-rich side the light blue squares correspond to unbound isotopes predicted to have half-lives larger than $\sim 10^{-9}$~s.}
\label{f:year-hafnium}
\end{figure}

Figure \ref{f:year-hafnium} summarizes the year of first discovery for all hafnium isotopes identified by the method of discovery. The range of isotopes predicted to exist is indicated on the right side of the figure. The radioactive hafnium isotopes were produced using fusion-evaporation reactions (FE), neutron capture reactions (NC), light-particle reactions (LP), spallation (SP), projectile fragmentation or fission (PF), and heavy-ion transfer reactions (TR). The stable isotopes were identified using mass spectroscopy (MS). In the following, the discovery of each hafnium isotope is discussed in detail and a summary is presented in Table 1.

\subsection*{Stable isotopes $^{174,176-180}$Hf}

Dempster described the discovery of $^{174}$Hf in the 1939 paper ``Isotopic constitution of hafnium, yttrium, lutetium and tantalum'' \cite{1939Dem01}. Hafnium in the form of doubly charged ions was studied with the Chicago mass spectrometer. ``In addition to the five isotopes reported by Aston, at 180, 179, 178, 177, and 176, a new isotope was found at 174 on six photographs.''

$^{176}$Hf, $^{177}$Hf, $^{178}$Hf, $^{179}$Hf, and $^{180}$Hf were discovered by Aston in ``Constitution of hafnium and other elements'' in 1934 \cite{1934Ast02}. The stable isotopes were identified with an anode discharge tube installed at the Cavendish Laboratory mass spectrograph. ``Hafnium gives a mass-spectrum indicating five isotopes, a weak line at 176 and four strong ones, 177, 178, 179, 180, of which the even numbers are rather more abundant.''

\subsection*{$^{154,155}$Hf}
In the 1981 paper ``New neutron deficient isotopes in the range of elements Tm to Pt'' Hofmann et al.\ reported the first observation of $^{154}$Hf and $^{155}$Hf \cite{1981Hof01}. Neutron deficient isotopes of elements from molybdenum to tin and vanadium to nickel targets were bombarded with $^{58}$Ni and $^{107}$Ag at the GSI linear accelerator UNILAC. Reaction products were separated by the SHIP velocity filter and implanted into silicon detectors. ``The time distances between parent and daughter of the 5 correlated events are between 2.0~s and 4.5~s. We explain these observations by the decay chain $^{158}$W $\xrightarrow{\alpha} ^{154}$Hf $\xrightarrow{\beta} ^{154}$Lu $\xrightarrow{\beta} ^{154}$Yb $\xrightarrow{\alpha} ^{150}$Er. A half-life of (2$\pm$1)~s can be deduced for the isotope $^{154}$Hf... A correlation showed as daughter activity an $\alpha$ line at 5656 keV decaying with T$_{1/2}$ = 890~ms. At this energy no other $\alpha$ line is known in this area except that of $^{155}$Lu with T$_{1/2}$ = 70~ms. This is short compared to the measured 890~ms decay. Therefore, our observations can easily be described within the frame of the decay chain $^{159}$W $\xrightarrow{\alpha} ^{155}$Hf $\xrightarrow{\beta} ^{155}$Lu $\xrightarrow{\alpha} ^{151}$Tm.'' These  half-lives correspond to the presently adopted values for $^{154}$Hf and $^{155}$Hf.

\subsection*{$^{156}$Hf}
``Alpha decay studies of very neutron deficient isotopes of Hf, Ta, W, and Re'' was published in 1979 by Hofmann et al.\ describing the observation of $^{156}$Hf \cite{1979Hof01}. Targets of $^{103}$Rh, $^{nat,108,110}$Pd, and $^{107,109}$Ag were bombarded with beams of $^{58}$Ni from the GSI UNILAC linear accelerator. Evaporation residues were separated with the high-velocity SHIP separator. ``The correlation method gave a half life of (25$\pm$4)~ms for $^{156}$Hf and an alpha branching ratio of 100\% ($>$81\%).''  This half-life agrees with the currently accepted value of 23(1)~ms. Previously only an upper limit of 30~ms was reported for the half-life of $^{156}$Hf \cite{1965Mac01}.

\subsection*{$^{157,158}$Hf}
Macfarlane discovered $^{157}$Hf and $^{158}$Hf in 1963 as reported in ``Alpha-decay properties of some lutetium and hafnium isotopes near the 82-neutron closed shell'' \cite{1965Mac01}. An enriched $^{144}$Sm target was bombarded with $^{20}$Ne at the Berkeley heavy-ion linear accelerator and $^{157}$Hf and $^{158}$Hf were produced in (7n) and (6n) fusion evaporation reactions, respectively. Excitation functions and $\alpha$-particles spectra were measured. ``One of the Hf alpha groups was observed at an alpha particle energy of 5.68 MeV and was found to decay with a half-life of 0.12$\pm$0.03~sec. This group can be seen in the spectra shown in [the figures]. The excitation function for the production of this activity by the reaction Sm$^{144}$+Ne$^{20}$, has a peak cross section of 0.2~mb at an excitation energy of 106~MeV. This energy agrees closely with values observed for (H.I.,7n) reactions, which means that this activity is due to Hf$^{157}$. The second Hf alpha group has an alpha-particle energy of 5.27~MeV and decays with a half-life of 3$\pm$0.5~sec... The results obtained above, therefore, are consistent with the mass assignment of Hf$^{158}$ to this activity.'' These half-lives agree with the presently adopted values of 115(1)~ms and 2.84(7)~s for $^{157}$Hf and $^{158}$Hf, respectively.

\subsection*{$^{159-161}$Hf}
The first observation of $^{159}$Hf, $^{160}$Hf, and $^{161}$Hf were reported in the 1973 paper ``Study of hafnium $\alpha$ emitters: New isotopes $^{159}$Hf, $^{160}$Hf, and $^{161}$Hf'' by Toth et al.\ \cite{1973Tot01}. Enriched $^{144}$Sm and $^{147}$Sm targets were bombarded with 80$-$153~MeV and 121$-$153~MeV $^{20}$Ne beams from the Oak Ridge isochronous cyclotron, respectively. Recoil products were swept into a collecting surface with helium gas where $\alpha$ particles were measured with a Si(Au) detector. ``The decay characteristics and mass assignments (made on the basis of yield curve measurements, cross bombardments, and parent-daughter relationships) of the three new $\alpha$ emitters are as follows: (1) $^{159}$Hf, E$_\alpha$ = 5.09$\pm$0.01~MeV, T$_{1/2}$ = 5.6$\pm$0.5~sec; (2) $^{160}$Hf, E$_\alpha$=4.77$\pm$0.02~MeV, T$_{1/2} \sim$ 12~sec; and (3) $^{161}$Hf, E$_\alpha$ = 4.60$\pm$0.01~MeV, T$_{1/2}$ = 17$\pm$2~sec.'' These half-lives agree with the currently adopted values of 5.20(10)~s, 13.6(2)~s, and 18.2(5)~s, respectively.

\subsection*{$^{162,163}$Hf}
In 1982 $^{162}$Hf and $^{163}$Hf were discovered by Schrewe et al.\ in ``Decay studies of the new isotopes $^{162,163}$Hf'' \cite{1982Sch01}. $^{24}$Mg beams accelerated to 105$-$133~MeV by the Chalk River MP tandem accelerator bombarded enriched $^{142}$Nd targets and $^{162}$Hf and $^{163}$Hf were produced in (4n) and (3n) evaporation reactions, respectively. Recoils were transported to a measuring station with a He-jet and $\beta$-delayed $\gamma$-spectra were measured with intrinsic Ge and Ge(Li) detectors. ``The 71~keV line was therefore assigned to the decay of $^{163}$Hf, the 174~keV line to the decay of $^{162}$Hf. In addition to these two lines, further $\gamma$-ray lines from $^{163,162}$Hf decays are summarized in [the tables]. Half-life determination was possible for most of these $\gamma$ rays, and yielded a mean half-life of T$_{1/2}$ = (37.6$\pm$0.8)~s for $^{162}$Hf and T$_{1/2}$ = (40.0$\pm$0.6)~s for $^{163}$Hf.'' The half-life for $^{162}$Hf is included in the calculation of the presently adopted value and the half-life of $^{163}$Hf corresponds to the current value.

\subsection*{$^{164,165}$Hf}
Bruchertseifer and Eichler reported the observation of $^{164}$Hf and $^{165}$Hf in the 1981 paper ``Untersuchung der Produkte der Reaktion $^{147}$Sm + $^{22}$Ne'' \cite{1981Bru01}.  Enriched $^{147}$Sm targets were bombarded with 110~MeV and 136~MeV $^{22}$Ne beams from the Dubna U300 accelerator and $^{164}$Hf and $^{165}$Hf were produced in (5n) and (4n) fusion-evaporation reactions, respectively. X-ray and $\gamma$-ray spectra were measured following chemical separation. ``Das Auftreten von $^{165}$Lu sowie $^{164}$Lu neben bekannten Hafniumisotopen nach der hocheffektiven Abtrennung (Trenneffekt $\ge$ 10$^3$) der Lanthanide belegt, dass die Hafniumfraktion die Isotope $^{164}$Hf und $^{165}$Hf enth\"ahlt.'' [The appearance of $^{165}$Lu as well as $^{164}$Lu next to known hafnium isotopes following the highly effective separation (separation efficiency $\ge$ 10$^3$) of the lanthanides proves that the hafnium fraction contains the isotopes $^{164}$Hf and $^{165}$Hf.] The reported half-lives of 2.8(2)~min ($^{164}$Hf) and 1.7(1)~min ($^{165}$Hf) are within a factor of two of the presently adopted values of 111(8)~s and 76(4)~s, respectively.

\subsection*{$^{166}$Hf}
In the 1965 paper ``Rotational states produced in heavy-ion nuclear reactions'' Stephens et al.\ described the observation of $^{166}$Hf \cite{1965Ste01}. The Berkeley Hilac accelerator was used to bombard self-supporting $^{159}$Tb targets with 117 MeV $^{14}$N beams and $^{166}$Hf was populated in the (7n) evaporation reaction. Conversion electron spectra were measured with a single wedge-gap electron spectrometer and $\gamma$-ray spectra with NaI(Tl) and Ge detectors. ``The de-excitation of the ground state rotational band in nine deformed even nuclei has been observed following heavy-ion nuclear reactions. The transitions from the states up to spin 16 (on the average) were observed and their energies were measured with an accuracy of $\pm$0.3\%.'' The rotational band in $^{166}$Hf was measured up to spin 14. Four years later Arlt et al.\ reported the first observation of the $^{166}$Hf ground state \cite{1969Arl04}.

\subsection*{$^{167}$Hf}
$^{167}$Hf was observed by Arlt et al.\ as reported in the 1969 paper ``The new neutron-deficient isotopes $^{169}$Hf, $^{167}$Hf, $^{166}$Hf, and $^{166}$Lu and the decay scheme of $^{169}$Hf'' \cite{1969Arl04}. Protons were accelerated to 660~MeV by the Dubna JINR synchrocyclotron and bombarded Ta$_2$O$_5$ targets to form hafnium isotopes in the Ta(p,2pxn) reaction which subsequently decayed to lutetium isotopes. Gamma-ray spectra were measured with NaI(Tl) and Ge(Li) detectors in singles and coincidences following chemical separation. ``The half-life was evaluated more accurately by extrapolating the decay curves isolated after equal accumulation intervals. The best conditions were provided by a 3-min irradiation, an accumulation time of 2 min, and separation of four Lu sources. The resulting half-life for $^{167}$Hf was 1.9$\pm$0.2~min.'' This half-life agrees with the currently accepted value of 2.05(5)~min. A previously reported half-life of $\sim$10~min \cite{1961Mer02} was evidently incorrect.

\subsection*{$^{168}$Hf}
The first observation of $^{168}$Hf was described by Merz and Caretto in ``Neutron-deficient nuclides of hafnium and lutetium'' in 1961 \cite{1961Mer02}. Tantalum, tungsten and Lu$_2$O$_3$ targets were irradiated with 300-400 MeV protons from the Carnegie Institute of Technology synchrocyclotron. After chemical separation $\gamma$-rays and positrons were measured with a NaI crystal and a magnetic spectrometer with an anthracene crystal, respectively. ``The third hafnium isotope reported Hf$^{168}$ has a half-life of (22$\pm$2) min, measured directly and indirectly by milking its 7.0-min lutetium daughter at intervals of 30 min.'' This half-life agrees with the presently adopted value of 25.95(20)~min.

\subsection*{$^{169}$Hf}
$^{169}$Hf was observed by Arlt et al.\ as reported in the 1969 paper ``The new neutron-deficient isotopes $^{169}$Hf, $^{167}$Hf, $^{166}$Hf, and $^{166}$Lu and the decay scheme of $^{169}$Hf'' \cite{1969Arl04}. Protons were accelerated to 660~MeV by the Dubna JINR synchrocyclotron and bombarded Ta$_2$O$_5$ targets to form hafnium isotopes in the Ta(p,2pxn) reaction. Gamma-ray spectra were measured with NaI(Tl) and Ge(Li) detectors in singles and coincidences following chemical separation. ``To clear up the discrepancies in the data on the A = 169 chain we examined the accumulation of $^{169}$Lu after 6 min in the Hf fraction isolated from a Ta target 4 min after the end of the 20 min irradiation. [The Figure] shows the decay curve of the parent as plotted from the intensities of the 191 and 96.2 keV $\gamma$ rays of $^{169}$Lu, and the 198 keV $\gamma$ ray of $^{169}$Yb in Lu preparations separated at 6 min from the Hf fraction. Three runs gave T = 5 $\pm$ 0.5~min for the parent isobar. We have assigned this half-life to $^{169}$Hf on the assumption that the Hf fraction did not contain any other elements.'' This half-life is close to the presently adopted value of 3.24(4)~min. A previously reported half-life of 1.5~h \cite{1961Mer02} was evidently incorrect. An unsuccessful search for the 1.5~h half-life determined an upper limit of 8~min \cite{1966Har01}.

\subsection*{$^{170}$Hf}
The first identification of $^{170}$Hf was described by Merz and Caretto in ``Neutron-deficient nuclides of hafnium and lutetium'' in 1961 \cite{1961Mer02}. Tantalum, tungsten and Lu$_2$O$_3$ targets were irradiated with 300-400 MeV protons from the Carnegie Institute of Technology synchrocyclotron. After chemical separation $\gamma$-rays and positrons were measured with a NaI crystal and a magnetic spectrometer with an anthracene crystal, respectively. ``A previously unreported activity with a half-life of (9$\pm$2)~hr was observed in the hafnium fraction, which cannot result from the reported 12- or 16-hr activities for Hf$^{171}$, because the half-life reported in this work was derived from milking a 1.9-day lutetium activity from the hafnium fraction. The 1.9-day lutetium activity is Lu$^{170}$.'' This half-life is within a factor of two of the presently adopted value of 16.01(13)~h.

\subsection*{$^{171-173}$Hf}
In ``Radioactive isotopes of lutetium and hafnium'' Wilkinson and Hicks described the identification of $^{171}$Hf, $^{172}$Hf, and $^{173}$Hf in 1951 \cite{1951Wil01}. Ytterbium was irradiated with 20 and 38~MeV $\alpha$-particles from the Berkeley 60-in.\ cyclotron and lutetium was irradiated with 15 to 75~MeV protons from the linear accelerator. Decay curves, absorption curves, and electron spectra were measured following chemical separation. ``16.0$\pm$0.5 hour Hf$^{171}$ --- Successive chemical separations of lutetium showed that the decay of the 16-hour activity formed the 8.5-day Lu$^{171}$... $\sim$5 year Hf$^{171}$ [This is apparently a typo and should read Hf$^{172}$] --- The decay has been followed for over two years, and the half-life appears to be about five years. The activity is the parent of a 6.70-day lutetium daughter; both the decay of the separated daughter and its growth in purified hafnium have been studied... 23.6$\pm$0.1 hour Hf$^{173}$ --- In the bombardments of ytterbium with 20-Mev alpha-particles and of lutetium with 18- to 32-Mev protons from the linear accelerator, an activity of 23.6 hours half-life was obtained; the radiation characteristics were identical in both cases.'' While the $^{171}$Hf half-life agrees with the currently adopted value of 12.1(4)~h, the $^{172}$Hf was not well determined and differs from the accepted value of 1.87(3)~y; these observations were accepted because of the correct identification of the daughter activities. The observed half-life for $^{173}$Hf of 23.6(1)~h corresponds to the presently adopted value.


\subsection*{$^{175}$Hf}
The first observation of $^{175}$Hf was reported by Wilkinson and Hicks in 1949 in ``Hf$^{175}$, a new radioactive isotope of hafnium'' \cite{1949Wil02}. Lutetium oxide was bombarded with 19 MeV deuterons and 10 MeV protons from the Berkeley 60-in.\ cyclotron. Decay curves and absorption spectra were measured with mica window counters following chemical separations. ``The chemically separated hafnium has been found to contain a single radioactivity, emitting electrons and $\gamma$-radiations, which decays with a half-life of 70$\pm$2 days... The new isotope is allocated to mass 175.'' This half-life corresponds to the currently adopted value.


\subsection*{$^{181}$Hf}
In 1935 Hevesy identified $^{181}$Hf in ``Radiopotassium and other artificial radio-elements'' \cite{1935Hev02}. Hafnium was irradiated with neutrons from a beryllium-radium source and $^{181}$Hf was formed in the reaction $^{180}$Hf(n,$\gamma$). Beta-rays were measured following chemical separation. ``The disintegration of the active hafnium is much slower than that of zirconium, half of the activity acquired being lost only after the lapse of a few months.'' Hevesy published a more accurate value of 55(10)~d later \cite{1938Hev01}. This half-life is consistent with the currently accepted value of 42.39(6)~d.

\subsection*{$^{182}$Hf}
In the 1961 paper ``The nuclide $^{182}$Hf'', Hutchin and Lindner reported the first observation of $^{182}$Hf \cite{1961Hut01}. A hafnum oxide target was irradiated with thermal neutron at the Idaho Falls Materials Testing Reactor. Gamma-ray spectra were recorded with a Na(Tl) scintillation detector and mass spectra were measured following chemical separation. ``A small specimen of the hafnium was analysed in the two-stage mass spectrometer at the Vallecitos Laboratory of the General Electric Company and found to contain 0.0088 atom per cent abundance at the mass 182 position. A small background correction was applied by a similar analysis of unirradiated hafnium. On the basis of the atomic abundance and the specific activity of the sample, a half-life of 8.5$\times$10$^6$ years was calculated for $^{181}$Hf.'' This half-life agrees with the currently accepted value of 9(2)~My. Two weeks later the observation of $^{182}$Hf was independently reported by Naumann and Michel \cite{1961Nau01}.

\subsection*{$^{183}$Hf}
$^{183}$Hf was discovered by Gatti and Flegenheimer in ``Ein neues Hf-Isotop (Hf-183)'' in 1956 \cite{1956Gat01}. Tungsten targets were irradiated with fast neutrons which were produced by bombarding beryllium targets with 28~MeV deuterons from the Buenos Aires synchrocyclotron. Decay curves, absorption and $\beta$-ray spectra were measured following chemical separation. ``Auf Grund der Kernreaktion, die zur Bildung des 64-min-Hf f\"uhrt (n,$\alpha$) und seinem Q$^{\beta^-}$-Wert von 2.2 MeV wird die Massenzahl 183 f\"ur das 64-min-Hf vorgeschlagen.'' [Because of the nuclear reaction that leads to the formation of the 64-min Hf (n,$\alpha$) and its Q$^{\beta^-}$ value, the mass number 183 is recommended for this 64-min Hf.] This half-life agrees with the presently adopted value of 1.067(17)~h.

\subsection*{$^{184}$Hf}
Ward et al.\ reported the first observation of $^{184}$Hf in the 1973 paper ``Identification of $^{184}$Hf'' \cite{1973War01}. Natural tungsten and enriched $^{186}$WO$_3$ targets were bombarded with 92 MeV and 200 MeV protons, respectively, and $^{184}$Hf was produced in the reaction $^{186}$W(p,3p). Gamma- and beta-ray spectra were measured following chemical and mass separation. ``Well-known $\gamma$-ray lines due to $^{184}$Ta can be seen in [the figure]. These were observed to grow and decay as expected for a 4.12-h parent feeding an 8.55-h daughter. Lines at 139.1, 181.0, and 344.9 keV decayed with a half-life of 4.12$\pm$0.05~h and we assign these $\gamma$ rays to $^{184}$Hf.'' This half-life of 4.12(5)~h corresponds to the presently adopted value. A previously reported 2.2(1)~h half-life was only reported in an unpublished report \cite{1961Mer03} which subsequently was referred to by a cross section measurement of the reactions $^{186}$W(p,3p) and $^{187}$Re(p,4p) \cite{1962Mor01}. This latter reference did not contain any information about properties of $^{184}$Hf.

\subsection*{$^{185}$Hf}
$^{185}$Hf was discovered by Yuan et al.\ in ``New neutron-rich nuclide $^{185}$Hf'' in 1993 \cite{1993Yua02}. Natural tungsten targets were irradiated with 14 MeV neutrons produced by bombarding a TiT target with deuterons from the Lanzhou 300-KV Cockroft-Walton accelerator. $^{185}$Hf was produced in the reaction $^{186}$W(n,2p) and identified by measuring $\gamma$-ray spectra with a HPGe detector following chemical separation. ``A radioactive-series decay analyzing program was applied resulting in the half-lives of 3.5$\pm$0.6~min and 48.6$\pm$5.6~min, for $^{185}$Hf and $^{185}$Ta, respectively.'' This half-life corresponds to the currently accepted value.

\subsection*{$^{186}$Hf}
In the 1998 paper ``Production and identification of a new heavy neutron-rich isotope $^{186}$Hf'' Yuan et al.\ reported the observation of $^{186}$Hf \cite{1998Yua01}. A 60~MeV/nucleon $^{18}$O beam bombarded a natural tungsten target and $^{186}$Hf was produced in multi-nucleon transfer reactions. Gamma-ray spectra were measured with a GMX HPGe detector following chemical separation. ``The assignment of the new nuclide $^{186}$Hf was primarily based on the time variation of the $\gamma$ rays of its daughter $^{186}$Ta.'' The reported half-life of 2.6(12)~min corresponds to the currently adopted value.

\subsection*{$^{187,188}$Hf}
Benlliure et al. published the discovery of $^{187}$Hf and $^{188}$Hf in the 1999 paper entitled ``Production of neutron-rich isotopes by cold fragmentation in the reaction $^{197}$Au + Be at 950 A MeV'' \cite{1999Ben01}. A 950~A$\cdot$MeV $^{197}$Au beam from the SIS synchrotron of GSI was incident on a beryllium target and $^{187}$Hf and $^{188}$Hf were produced in projectile fragmentation reactions. The FRS fragment separator was used to select isotopes with a specific mass-to-charge ratio. ``In the right part of [the figure] the projected A/Z distributions are shown for the different elements transmitted in this setting of the FRS. In this setting the isotopes $^{193}$Re, $^{194}$Re, $^{191}$W, $^{192}$W, $^{189}$Ta, $^{187}$Hf and $^{188}$Hf were clearly identified for the first time. Only isotopes with a yield higher than 15 counts were considered as unambiguously identified.''

\subsection*{$^{189}$Hf}
Alkhomashi et al. observed $^{189}$Hf in the 2009 paper ``$\beta^-$-delayed spectroscopy of neutron-rich tantalum nuclei: Shape evolution in neutron-rich tungsten isotopes'' \cite{2009Alk01}. A beryllium target was bombarded with a 1 GeV/nucleon $^{208}$Pb beam from the SIS-18 heavy-ion synchrotron at GSI, Germany. Projectile-like fragments were separated with the FRS and implanted in a series of double-sided silicon strip detectors where correlated $\beta$-decay was measured in coincidence with $\gamma$-rays in the $\gamma$-ray spectrometer RISING. Although not specifically mentioned in the text, evidence for $^{189}$Hf is clearly visible in the two-dimensional particle identification plot.

\section{Summary}

The discoveries of the known gallium, germanium, lutetium, and hafnium isotopes have been compiled and the methods of their production discussed.

While in gallium only $^{64}$Ga and $^{74}$Ga were at first incorrectly identified, five germanium isotopes were initially incorrectly assigned ($^{65}$Ge, $^{67}$Ge, $^{69}$Ge, $^{71}$Ge, and $^{78}$Ge). In addition, the half-lives of $^{70}$Ga and $^{72}$Ga were first observed without a mass assignment and $^{71}$Ge, $^{75}$Ge, and $^{77}$Ge were incorrectly reported to be stable.

The discovery of the two heavier elements was fairly straightforward; only $^{177}$Lu, $^{180}$Lu, $^{167}$Hf, and $^{169}$Hf were initially identified incorrectly and the half-lives of the four lutetium isotopes $^{177}$Lu, $^{178}$Lu, $^{179}$Lu, and $^{180}$Lu were at first reported without a mass assignment.

\ack

This work was supported by the National Science Foundation under grant No. PHY06-06007.

\bibliography{../isotope-discovery-references}

\newpage

\newpage

\TableExplanation

\bigskip
\renewcommand{\arraystretch}{1.0}

\section{Table 1.\label{tbl1te} Discovery of gallium, germanium, lutetium, and hafnium isotopes }
\begin{tabular*}{0.95\textwidth}{@{}@{\extracolsep{\fill}}lp{5.5in}@{}}
\multicolumn{2}{p{0.95\textwidth}}{ }\\

Isotope & Gallium, germanium, lutetium, and hafnium isotope \\
First author & First author of refereed publication \\
Journal & Journal of publication \\
Ref. & Reference \\
Method & Production method used in the discovery: \\

  & FE: fusion evaporation \\
  & NC: Neutron capture reactions \\
  & PN: photo-nuclear reactions \\
  & LP: light-particle reactions (including neutrons) \\
  & MS: mass spectroscopy \\
  & TR: heavy-ion transfer reactions \\
  & NF: neutron induced fission \\
  & DI: deep inelastic reaction \\
  & SP: spallation reactions \\
  & PF: projectile fragmentation of fission \\

Laboratory & Laboratory where the experiment was performed\\
Country & Country of laboratory\\
Year & Year of discovery \\
\end{tabular*}
\label{tableI}

\datatables 



\setlength{\LTleft}{0pt}
\setlength{\LTright}{0pt}


\setlength{\tabcolsep}{0.5\tabcolsep}

\renewcommand{\arraystretch}{1.0}

\footnotesize 

\begin{longtable}{@{\extracolsep\fill}llllllll@{}}
\caption{Discovery of gallium, germanium, lutetium, and hafnium isotopes. See page\ \pageref{tbl1te} for Explanation of Tables}
Isotope & First Author & Journal & Ref. & Method & Laboratory & Country & Year\\
\hline\\
\endfirsthead\\
\caption[]{(continued)}
Isotope & First author & Journal & Ref. & Method & Laboratory & Country & Year\\
\hline\\
\endhead
$^{60}$Ga & B. Blank & Phys. Rev. Lett. &\cite{1995Bla01}& PF & GANIL & France &1995 \\
$^{61}$Ga & M.A.C. Hotchkins & Phys. Rev. C &\cite{1987Hot01}& FE & Berkeley & USA &1987 \\
$^{62}$Ga & R. Chiba & Phys. Rev. C &\cite{1978Chi01}& LP & Tokyo & Japan &1978 \\
$^{63}$Ga & M. Nurmia & Phys. Lett. &\cite{1965Nur01}& FE & Argonne & USA &1965 \\
$^{64}$Ga & B.L. Cohen & Phys. Rev. &\cite{1953Coh01}& LP & Oak Ridge & USA &1953 \\
$^{65}$Ga & L.W. Alvarez & Phys. Rev. &\cite{1938Alv02}& LP & Berkeley & USA &1938 \\
$^{66}$Ga & W.B. Mann & Phys. Rev. &\cite{1937Man01}& LP & Berkeley & USA &1937 \\
$^{67}$Ga & L.W. Alvarez & Phys. Rev. &\cite{1938Alv01}& LP & Berkeley & USA &1938 \\
$^{68}$Ga & W. Bothe & Naturwiss. &\cite{1937Bot03}& PN & Heidelberg & Germany &1937 \\
$^{69}$Ga & F.W. Aston & Nature &\cite{1923Ast01}& MS & Cambridge & UK &1923 \\
$^{70}$Ga & W. Bothe & Naturwiss. &\cite{1937Bot03}& PN & Heidelberg & Germany &1937 \\
$^{71}$Ga & F.W. Aston & Nature &\cite{1923Ast01}& MS & Cambridge & UK &1923 \\
$^{72}$Ga & R. Sagane & Phys. Rev. &\cite{1939Sag01}& NC & Berkeley & USA &1939 \\
$^{73}$Ga & M.L. Perlman & Phys. Rev. &\cite{1949Per01}& PN & General Electric Research Laboratory& USA &1949 \\
$^{74}$Ga & H. Morinaga & Phys. Rev. &\cite{1956Mor01}& LP & Purdue & USA &1956 \\
$^{75}$Ga & H. Morinaga & J. Phys. Soc. Japan &\cite{1960Mor01}& PN & Tohoku & Japan &1960 \\
$^{76}$Ga & K. Takahashi & J. Phys. Soc. Japan &\cite{1961Tak01}& LP & Tokyo & Japan &1961 \\
$^{77}$Ga & L. Wish & Phys. Rev. &\cite{1968Wis01}& NF & Naval Radiological Defense Laboratory & USA &1968 \\
$^{78}$Ga & P. del Marmol & Nucl. Phys. A &\cite{1972del01}& NF & Mol & Belgium &1972 \\
$^{79}$Ga & B. Grapengiesser & J. Inorg. Nucl. Chem. &\cite{1974Gra01}& NF & Studsvik & Sweden &1974 \\
$^{80}$Ga & B. Grapengiesser & J. Inorg. Nucl. Chem. &\cite{1974Gra01}& NF & Studsvik & Sweden &1974 \\
$^{81}$Ga & G. Rudstam & Phys. Rev. C &\cite{1976Rud01}& NF & Studsvik & Sweden &1976 \\
$^{82}$Ga & G. Rudstam & Phys. Rev. C &\cite{1976Rud01}& NF & Studsvik & Sweden &1976 \\
$^{83}$Ga & G. Rudstam & Phys. Rev. C &\cite{1976Rud01}& NF & Studsvik & Sweden &1976 \\
$^{84}$Ga & K.-L. Kratz & Z. Phys. A &\cite{1991Kra01}& SP & CERN & Switzerland &1991 \\
$^{85}$Ga & M. Bernas & Phys. Lett. B &\cite{1997Ber01}& PF & Darmstadt & Germany &1997 \\
$^{86}$Ga & M. Bernas & Phys. Lett. B &\cite{1997Ber01}& PF & Darmstadt & Germany &1997 \\
$^{87}$Ga & T. Ohnishi & J. Phys. Soc. Japan &\cite{2010Ohn01}& PF & RIKEN & Japan &2010 \\
 & & & & & &  \\
 & & & & & &  \\
$^{60}$Ge & A. Stolz & Phys. Lett. B &\cite{2005Sto01}& PF & Michigan State & USA &2005 \\
$^{61}$Ge & M.A.C. Hotchkins & Phys. Rev. C &\cite{1987Hot01}& FE & Berkeley & USA &1987 \\
$^{62}$Ge & M.F. Mohar & Phys. Rev. Lett. &\cite{1991Moh01}& PF & Michigan State & USA &1991 \\
$^{63}$Ge & M.F. Mohar & Phys. Rev. Lett. &\cite{1991Moh01}& PF & Michigan State & USA &1991 \\
$^{64}$Ge & R.G.H. Robertson & Phys. Rev. Lett. &\cite{1972Rob01}& LP & Michigan State & USA &1972 \\
$^{65}$Ge & R.G.H. Robertson & Phys. Rev. Lett. &\cite{1972Rob01}& LP & Michigan State & USA &1972 \\
$^{66}$Ge & H.H. Hopkins Jr. & Phys. Rev. &\cite{1950Hop01}& LP & Berkeley & USA &1950 \\
$^{67}$Ge & H.H. Hopkins Jr. & Phys. Rev. &\cite{1950Hop01}& LP & Berkeley & USA &1950 \\
$^{68}$Ge & H.H. Hopkins Jr. & Phys. Rev. &\cite{1948Hop01}& LP & Berkeley & USA &1948 \\
$^{69}$Ge & W.B. Mann & Phys. Rev. &\cite{1938Man01}& LP & Berkeley & USA &1938 \\
$^{70}$Ge & F.W. Aston & Nature &\cite{1923Ast03}& MS & Cambridge & UK &1923 \\
$^{71}$Ge & G.T. Seaborg & Phys. Rev. &\cite{1941Sea01}& LP & Berkeley & USA &1941 \\
$^{72}$Ge & F.W. Aston & Nature &\cite{1923Ast03}& MS & Cambridge & UK &1923 \\
$^{73}$Ge & K.T. Bainbridge & J. Frank. Inst. &\cite{1933Bai01}& MS & Bartol Research Foundation & USA &1933 \\
$^{74}$Ge & F.W. Aston & Nature &\cite{1923Ast03}& MS & Cambridge & UK &1923 \\
$^{75}$Ge & R. Sagane & Phys. Rev. &\cite{1939Sag01}& LP & Berkeley & USA &1939 \\
$^{76}$Ge & K.T. Bainbridge & J. Frank. Inst. &\cite{1933Bai01}& MS & Bartol Research Foundation & USA &1933 \\
$^{77}$Ge & R. Sagane & Phys. Rev. &\cite{1939Sag01}& LP & Berkeley & USA &1939 \\
$^{78}$Ge & N. Sugarman & Phys. Rev. &\cite{1953Sug01}& NF & Los Alamos & USA &1953 \\
$^{79}$Ge & M. Karras & Nucl. Phys. A &\cite{1970Kar01}& LP & Arkansas & USA &1970 \\
$^{80}$Ge & P. del Marmol & Nucl. Phys. A &\cite{1972del01}& NF & Mol & Belgium &1972 \\
$^{81}$Ge & P. del Marmol & Nucl. Phys. A &\cite{1972del01}& NF & Mol & Belgium &1972 \\
$^{82}$Ge & P. del Marmol & Nucl. Phys. A &\cite{1972del01}& NF & Mol & Belgium &1972 \\
$^{83}$Ge & P. del Marmol & Nucl. Phys. A &\cite{1972del01}& NF & Mol & Belgium &1972 \\
$^{84}$Ge & P. del Marmol & Nucl. Phys. A &\cite{1972del01}& NF & Mol & Belgium &1972 \\
$^{85}$Ge & J.P. Omtvedt & Z. Phys. A &\cite{1991Omt01}& NF & Studsvik & Sweden &1991 \\
$^{86}$Ge & M. Bernas & Phys. Lett. B &\cite{1994Ber01}& PF & Darmstadt & Germany &1994 \\
$^{87}$Ge & M. Bernas & Phys. Lett. B &\cite{1997Ber01}& PF & Darmstadt & Germany &1997 \\
$^{88}$Ge & M. Bernas & Phys. Lett. B &\cite{1997Ber01}& PF & Darmstadt & Germany &1997 \\
$^{89}$Ge & M. Bernas & Phys. Lett. B &\cite{1997Ber01}& PF & Darmstadt & Germany &1997 \\
$^{90}$Ge & T. Ohnishi & J. Phys. Soc. Japan &\cite{2010Ohn01}& PF & RIKEN & Japan &2010 \\
 & & & & & &  \\
 & & & & & &  \\
$^{150}$Lu& P.J. Sellin & Phys. Rev. C &\cite{1993Sel02}& FE & Daresbury & UK &1993 \\
$^{151}$Lu & S. Hofmann & Z. Phys. A &\cite{1982Hof01}& FE & Darmstadt & Germany &1982 \\
$^{152}$Lu & K.S. Toth & Phys. Rev. C &\cite{1987Tot01}& FE & Berkeley & USA &1987 \\
$^{153}$Lu & J.M. Nitschke & Z. Phys. A &\cite{1989Nit01}& FE & Berkeley & USA &1989 \\
$^{154}$Lu & S. Hofmann & Z. Phys. A &\cite{1981Hof01}& FE & Darmstadt & Germany &1981 \\
$^{155}$Lu & R.D. Macfarlane & Phys. Rev. &\cite{1965Mac01}& FE & Berkeley & USA &1965 \\
$^{156}$Lu & R.D. Macfarlane & Phys. Rev. &\cite{1965Mac01}& FE & Berkeley & USA &1965 \\
$^{157}$Lu & E. Hagberg & Nucl. Phys. A &\cite{1977Hag02}& SP & CERN & Switzerland &1977 \\
$^{158}$Lu & G.D. Alkhazov & Z. Phys. A &\cite{1979Alk01}& SP & Dubna & Russia &1979 \\
$^{159}$Lu & G.D. Alkhazov & Z. Phys. A &\cite{1980Alk01}& SP & Dubna & Russia &1980 \\
$^{160}$Lu & G.D. Alkhazov & Z. Phys. A &\cite{1979Alk01}& SP & Dubna & Russia &1979 \\
$^{161}$Lu & R. Anholt & Nucl. Phys. A &\cite{1973Anh01}& FE & Yale & USA &1973 \\
$^{162}$Lu & C. Burman & Can. J. Phys. &\cite{1978Bur01}& FE & Yale & USA &1978 \\
$^{163}$Lu & G.D. Alkhazov & Z. Phys. A &\cite{1979Alk01}& SP & Dubna & Russia &1979 \\
$^{164}$Lu & R.C. Hunter & Phys. Rev. C &\cite{1977Hun01}& FE & Oak Ridge & USA &1977 \\
$^{165}$Lu & B.J. Meijer & Radiochim. Acta &\cite{1973Mei01}& LP & Amsterdam & Netherlands &1973 \\
$^{166}$Lu & R. Arlt & Bull. Acad. Sci. USSR &\cite{1969Arl04}& SP & Dubna & Russia &1969 \\
$^{167}$Lu & P.M. Aron & Bull. Acad. Sci. USSR &\cite{1958Aro01}& SP & Dubna & Russia &1958 \\
$^{168}$Lu & R.G. Wilson & Phys. Rev. &\cite{1960Wil01}& LP & Ohio State & USA &1960 \\
$^{169}$Lu & W.E. Nervik & Phys. Rev. &\cite{1955Ner01}& SP & Berkeley & USA &1955 \\
$^{170}$Lu & G. Wilkinson & Phys. Rev. &\cite{1951Wil01}& LP & Berkeley & USA &1951 \\
$^{171}$Lu & G. Wilkinson & Phys. Rev. &\cite{1951Wil01}& LP & Berkeley & USA &1951 \\
$^{172}$Lu & G. Wilkinson & Phys. Rev. &\cite{1951Wil01}& LP & Berkeley & USA &1951 \\
$^{173}$Lu & G. Wilkinson & Phys. Rev. &\cite{1951Wil01}& LP & Berkeley & USA &1951 \\
$^{174}$Lu & G. Wilkinson & Phys. Rev. &\cite{1951Wil01}& LP & Berkeley & USA &1951 \\
$^{175}$Lu & F.W. Aston & Nature &\cite{1934Ast04}& MS & Cambridge & UK &1934 \\
$^{176}$Lu & J. K. Marsh & Nature &\cite{1935Mar01}& NC & Oxford & UK &1935 \\
$^{177}$Lu & H. Atterling & Arkiv Mat. Astron. Fysik&\cite{1945Att01}& NC & Stockholm & Sweden &1945 \\
$^{178}$Lu & T. Stribel & Z. Naturforsch. &\cite{1957Str01}& LP & Frankfurt & Germany &1957 \\
$^{179}$Lu & T. Kuroyanagi & J. Phys. Soc. Japan &\cite{1961Kur01}& PN & Tohoku & Japan &1961 \\
$^{180}$Lu & S.C. Gujrathi & Nucl. Phys. A &\cite{1971Guj01}& LP & Simon Fraser & Canada &1971 \\
$^{181}$Lu & R. Kirchner & Nucl. Phys. A &\cite{1982Kir01}& DI & Darmstadt & Germany &1982 \\
$^{182}$Lu & R. Kirchner & Nucl. Phys. A &\cite{1982Kir01}& DI & Darmstadt & Germany &1982 \\
$^{183}$Lu & K. Rykaczewski & Z. Phys. A &\cite{1983Ryk01}& DI & Darmstadt & Germany &1983 \\
$^{184}$Lu & K. Rykaczewski & Nucl. Phys. A &\cite{1989Ryk01}& DI & Darmstadt & Germany &1989 \\
 & & & & & &  \\
 & & & & & &  \\
$^{154}$Hf& S. Hofmann & Z. Phys. A &\cite{1981Hof01}& FE & Darmstadt & Germany &1981 \\
$^{155}$Hf & S. Hofmann & Z. Phys. A &\cite{1981Hof01}& FE & Darmstadt & Germany &1981 \\
$^{156}$Hf & S. Hofmann & Z. Phys. A &\cite{1979Hof01}& FE & Darmstadt & Germany &1979 \\
$^{157}$Hf & R.D. Macfarlane & Phys. Rev. &\cite{1965Mac01}& FE & Berkeley & USA &1965 \\
$^{158}$Hf & R.D. Macfarlane & Phys. Rev. &\cite{1965Mac01}& FE & Berkeley & USA &1965 \\
$^{159}$Hf & K.S. Toth & Phys. Rev. C &\cite{1973Tot01}& FE & Oak Ridge & USA &1973 \\
$^{160}$Hf & K.S. Toth & Phys. Rev. C &\cite{1973Tot01}& FE & Oak Ridge & USA &1973 \\
$^{161}$Hf & K.S. Toth & Phys. Rev. C &\cite{1973Tot01}& FE & Oak Ridge & USA &1973 \\
$^{162}$Hf & U.J. Schrewe & Phys. Rev. C &\cite{1982Sch01}& FE & Chalk River & Canada &1982 \\
$^{163}$Hf & U.J. Schrewe & Phys. Rev. C &\cite{1982Sch01}& FE & Chalk River & Canada &1982 \\
$^{164}$Hf & H. Bruchertseifer & Radiochem. Radioanal. Lett. &\cite{1981Bru01}& FE & Dubna & Russia &1981 \\
$^{165}$Hf & H. Bruchertseifer & Radiochem. Radioanal. Lett. &\cite{1981Bru01}& FE & Dubna & Russia &1981 \\
$^{166}$Hf & F.S. Stephens & Nucl. Phys. &\cite{1965Ste01}& FE & Berkeley & USA &1965 \\
$^{167}$Hf & R. Arlt & Bull. Acad. Sci. USSR &\cite{1969Arl04}& SP & Dubna & Russia &1969 \\
$^{168}$Hf & E.R. Merz & Phys. Rev. &\cite{1961Mer02}& SP & Pittsburgh & USA &1961 \\
$^{169}$Hf & R. Arlt & Bull. Acad. Sci. USSR &\cite{1969Arl04}& SP & Dubna & Russia &1969 \\
$^{170}$Hf & E.R. Merz & Phys. Rev. &\cite{1961Mer02}& SP & Pittsburgh & USA &1961 \\
$^{171}$Hf & G. Wilkinson & Phys. Rev. &\cite{1951Wil01}& LP & Berkeley & USA &1951 \\
$^{172}$Hf & G. Wilkinson & Phys. Rev. &\cite{1951Wil01}& LP & Berkeley & USA &1951 \\
$^{173}$Hf & G. Wilkinson & Phys. Rev. &\cite{1951Wil01}& LP & Berkeley & USA &1951 \\
$^{174}$Hf & A.J. Dempster & Phys. Rev. &\cite{1939Dem01}& MS & Chicago & USA &1939 \\
$^{175}$Hf & G. Wilkinson & Phys. Rev. &\cite{1949Wil02}& LP & Berkeley & USA &1949 \\
$^{176}$Hf & F.W. Aston & Nature &\cite{1934Ast02}& MS & Cambridge & UK &1934 \\
$^{177}$Hf & F.W. Aston & Nature &\cite{1934Ast02}& MS & Cambridge & UK &1934 \\
$^{178}$Hf & F.W. Aston & Nature &\cite{1934Ast02}& MS & Cambridge & UK &1934 \\
$^{179}$Hf & F.W. Aston & Nature &\cite{1934Ast02}& MS & Cambridge & UK &1934 \\
$^{180}$Hf & F.W. Aston & Nature &\cite{1934Ast02}& MS & Cambridge & UK &1934 \\
$^{181}$Hf & G. Hevesy & Nature &\cite{1935Hev02}& NC & Copenhagen & Denmark &1935 \\
$^{182}$Hf & W.H. Hutchin & J. Inorg. Nucl. Chem. &\cite{1961Hut01}& NC & Livermore & USA &1961 \\
$^{183}$Hf & O.O. Gatti& Z. Naturforsch. &\cite{1956Gat01}& LP & Buenos Aires & Argentina &1956 \\
$^{184}$Hf & T.E. Ward & Phys. Rev. C &\cite{1973War01}& LP & Brookhaven & USA &1973 \\
$^{185}$Hf & S. Yuan & Z. Phys. A &\cite{1993Yua02}& LP & Lanzhou & China &1993 \\
$^{186}$Hf & S. Yuan & Phys. Rev. C &\cite{1998Yua01}& TR & Lanzhou & China &1998 \\
$^{187}$Hf & J. Benlliure & Nucl. Phys. A &\cite{1999Ben01}& PF & Darmstadt & Germany &1999 \\
$^{188}$Hf & J. Benlliure & Nucl. Phys. A &\cite{1999Ben01}& PF & Darmstadt & Germany &1999 \\
$^{189}$Hf & N. Alkhomashi & Phys. Rev. C &\cite{2009Alk01}& PF & Darmstadt & Germany &2009 \\
\end{longtable}

\end{document}